\documentclass[sigconf]{acmart}
\usepackage[ruled,vlined]{algorithm2e}
\usepackage{multirow}
\usepackage{subfigure}
\usepackage{utfsym}

\AtBeginDocument{%
  \providecommand\BibTeX{{%
    \normalfont B\kern-0.5em{\scshape i\kern-0.25em b}\kern-0.8em\TeX}}}

\setcopyright{acmlicensed}

\copyrightyear{2024}
\acmYear{2024}
\setcopyright{acmlicensed}\acmConference[KDD '24]{Proceedings of the 30th ACM SIGKDD Conference on Knowledge Discovery and Data Mining}{August 25--29, 2024}{Barcelona, Spain}
\acmBooktitle{Proceedings of the 30th ACM SIGKDD Conference on Knowledge Discovery and Data Mining (KDD '24), August 25--29, 2024, Barcelona, Spain}
\acmDOI{10.1145/3637528.3671686}
\acmISBN{979-8-4007-0490-1/24/08}

\begin{document}

\title{Pre-train and Refine: Towards Higher Efficiency in K-Agnostic Community Detection without Quality Degradation}

\author{Meng Qin}
\email{mengqin\_az@foxmail.com}
\affiliation{%
  \institution{Department of CSE, HKUST}
  \country{Hong Kong SAR}
}
\author{Chaorui Zhang}
\email{chaorui.zhang@gmail.com}
\affiliation{%
  \institution{Theory Lab, Huawei}
  \country{Hong Kong SAR}
}
\author{Yu Gao}
\email{gaoyu99@huawei.com}
\affiliation{%
  \institution{Theory Lab, Huawei}
  \country{Beijing, China}
}
\author{Weixi Zhang}
\email{zhangweixi1@huawei.com}
\affiliation{%
  \institution{Theory Lab, Huawei}
  \country{Hong Kong SAR}
}
\author{Dit-Yan Yeung}
\email{dyyeung@cse.ust.hk}
\affiliation{%
  \institution{Department of CSE, HKUST}
  \country{Hong Kong SAR}
}

\renewcommand{\shortauthors}{Meng Qin, et al.}

\begin{abstract}
  Community detection (CD) is a classic graph inference task that partitions nodes of a graph into densely connected groups. While many CD methods have been proposed with either impressive quality or efficiency, balancing the two aspects remains a challenge. This study explores the potential of deep graph learning to achieve a better trade-off between the quality and efficiency of $K$-agnostic CD, where the number of communities $K$ is unknown. We propose PRoCD (\underline{P}re-training \& \underline{R}efinement f\underline{o}r \underline{C}ommunity \underline{D}etection), a simple yet effective method that reformulates $K$-agnostic CD as the binary node pair classification. PRoCD follows a \textit{pre-training \& refinement} paradigm inspired by recent advances in pre-training techniques. We first conduct the \textit{offline pre-training} of PRoCD on small synthetic graphs covering various topology properties. Based on the inductive inference across graphs, we then \textit{generalize} the pre-trained model (with frozen parameters) to large real graphs and use the derived CD results as the initialization of an existing efficient CD method (e.g., InfoMap) to further \textit{refine} the quality of CD results. In addition to benefiting from the transfer ability regarding quality, the \textit{online generalization} and \textit{refinement} can also help achieve high inference efficiency, since there is no time-consuming model optimization. Experiments on public datasets with various scales demonstrate that PRoCD can ensure higher efficiency in $K$-agnostic CD without significant quality degradation.
\end{abstract}





\maketitle

\section{Introduction}\label{Sec:Intro}
Community detection (CD) is a classic graph inference task that partitions nodes of a graph into several groups (i.e., communities) with dense linkage distinct from other groups \cite{fortunato202220}. It has been validated that the extracted communities may correspond to substructures of various real-world complex systems (e.g., functional groups in protein interactions \cite{fortunato2010community}).
Many network applications (e.g., cellular network decomposition \cite{dai2017optimal} and Internet traffic profiling \cite{qin2019towards}) are thus formulated as CD.
Due to the NP-hardness of some typical CD objectives (e.g., modularity maximization \cite{newman2006modularity}), balancing the quality and efficiency of CD on large graphs remains a challenge. Most existing approaches focus on either high quality or efficiency.

On the one hand, efficient CD methods usually adopt heuristic strategies or fast approximation w.r.t. some relaxed CD objectives to obtain feasible CD results (e.g., the greedy maximization \cite{blondel2008fast} and semi-definite relaxation \cite{wang2020community} of modularity).
Following the graph embedding framework, which maps nodes $\mathcal{V} = \{ v_i \}$ of a graph into low-dimensional vector representations $\{ {\bf{z}}_i \in \mathbb{R}^d \}$ ($d \ll |\mathcal{V}|$) with major topology properties preserved, several efficient embedding approaches have also been proposed based on approximated dimension reduction (e.g., random projection for high-order topology \cite{zhang2018billion}). Some of them are claimed to be community-preserving and able to support CD using a downstream clustering module \cite{bhowmick2020louvainne,gao2023raftgp}.
Despite the high efficiency, these methods may potentially suffer from quality degradation due to the information loss of heuristics, approximation, and relaxation.

On the other hand, recent studies have demonstrated the ability of deep graph learning (DGL) techniques \cite{rong2020deep,zhang2020deep}, e.g., those based on graph neural networks (GNNs), to achieve impressive quality of various graph inference tasks including CD \cite{xing2022comprehensive}. However, their powerful performance usually relies on iterative optimization algorithms (e.g., gradient descent) that direct sophisticated models to fit complicated objectives, which usually have high complexities.

\begin{figure}[t]
  \centering
  \includegraphics[width=\linewidth, trim=22 22 20 20,clip]{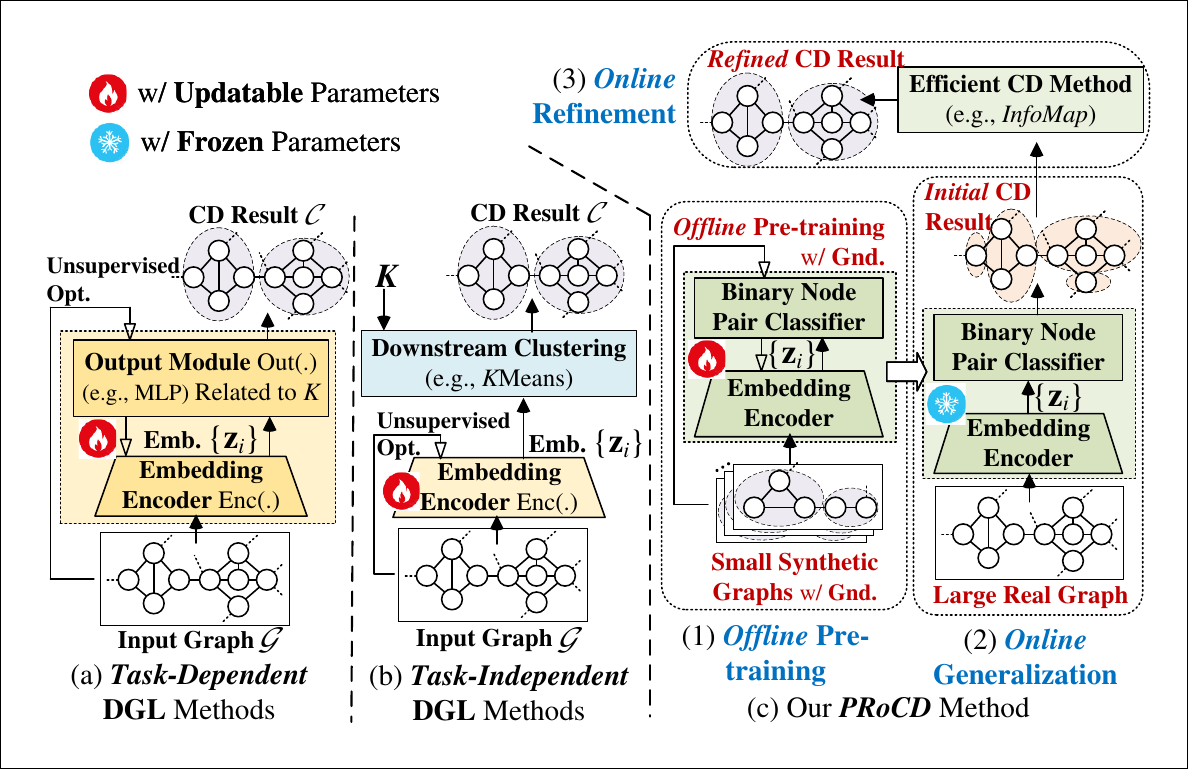}
  \caption{Overview of mainstream (a) \textit{task-dependent} and (b) \textit{-independent} DGL methods as well as (c) our PRoCD method.
  }\label{Fig:Overview}
\end{figure}

In this study, we explore the potential of DGL to ensure \textit{higher efficiency in CD without significant quality degradation}, compared with mainstream efficient and DGL methods. It serves as a possible way to achieve a better trade-off between the two aspects.
Different from existing CD methods \cite{dhillon2007weighted,chen2019supervised,tsitsulin2023graph} with an assumption that the number of communities $K$ is given, we consider the more challenging yet realistic $K$-agnostic CD, where $K$ is unknown. In this setting, one should simultaneously determine $K$ and the corresponding community partition.

\textbf{Dilemmas}.
As shown in Fig.~\ref{Fig:Overview} (a) and (b), existing DGL-based CD techniques usually follow the unified graph embedding framework and can be categorized into the (\romannumeral1) \textit{task-dependent} and (\romannumeral2) \textit{task-independent} approaches. We argue that they may suffer from the following limitations w.r.t. our settings.

First, \textit{existing DGL methods may be inapplicable to $K$-agnostic CD}. Given a graph $\mathcal{G}$, some \textit{task-dependent} approaches \cite{wilder2019end,chen2019supervised,tsitsulin2023graph} (see Fig.~\ref{Fig:Overview} (a)) generate embeddings $\{ {\bf{z}}_i \}$ via an embedding encoder ${\mathop{\rm Enc}\nolimits} ( \cdot )$ (e.g., a multi-layer GNN) and feed $\{ {\bf{z}}_i \}$ into an output module ${\mathop{\rm Out}\nolimits} ( \cdot )$ (e.g., a multi-layer perceptron (MLP)) to derive the CD result $\mathcal{C}$, i.e., $\{ {{\bf{z}}_i}\}  = {\mathop{\rm Enc}\nolimits} ( \mathcal{G} )$ and $\mathcal{C} = {\mathop{\rm Out}\nolimits} (\{ {{\bf{z}}_i}\} )$.
Since parameters in ${\mathop{\rm Out}\nolimits} ( \cdot )$ are usually with the dimensionality related to $K$, these \textit{task-dependent} methods cannot output a feasible CD result when $K$ is unknown.
Although some \textit{task-independent} methods \cite{yang2016modularity,liu2022deep,qin2023towards} (see Fig.~\ref{Fig:Overview} (b)) do not contain ${\mathop{\rm Out}\nolimits} ( \cdot )$ related to $K$, their original designs still rely on a downstream clustering algorithm (e.g., $K$Means) with $\{ {\bf{z}}_i \}$ and $K$ as required inputs.

Second, \textit{the standard transductive inference of some DGL techniques may result in low efficiency}. As illustrated in Fig.~\ref{Fig:Overview} (a) and (b), during the inference of CD on each single graph, these transductive methods must first optimize their model parameters from scratch via unsupervised objectives, which is usually time-consuming. Some related studies \cite{yang2016modularity,liu2022deep,tsitsulin2023graph} only consider \textit{transductive inference} regardless of its low efficiency.

Focusing on the CD with a given $K$ and \textit{task-independent} architecture in Fig.~\ref{Fig:Overview} (b), recent research \cite{qin2023towards} has validated that the \textit{inductive inference}, which (\romannumeral1) trains a DGL model on historical graphs $\{ {\mathcal{G}}_t \}$ and (\romannumeral2) generalizes it to new graphs $\{ \mathcal{G}' \}$ without additional optimization, can help achieve high inference efficiency on $\{ \mathcal{G}' \}$. One can extend this method to $K$-agnostic CD by replacing the downstream clustering module (e.g., $K$Means) with an advanced algorithm unrelated to $K$ (e.g., DBSCAN \cite{schubert2017dbscan}). Our experiments indicate that such a naive extension may still have low inference quality and efficiency compared with conventional baselines.

\textbf{Contributions}.
We propose PRoCD (\underline{P}re-training \& \underline{R}efinement f\underline{o}r \underline{C}ommunity \underline{D}etection), a simple yet effective method as illustrated in Fig.~\ref{Fig:Overview} (c), to address the aforementioned limitations.

\textit{\textbf{A novel model architecture}}.
To derive feasible results for $K$-agnostic CD in an end-to-end way, we reformulate CD as the binary node pair classification. Unlike existing end-to-end models that directly assign the community label $c_i \in \{ 1, \cdots, K\}$ of each node $v_i$ (e.g., via an MLP in Fig.~\ref{Fig:Overview} (a)), we develop a binary classifier, with a new design of \textit{pair-wise temperature parameters}, to determine whether a pair of nodes $(v_i, v_j)$ are partitioned into the same community. Given the classification result w.r.t. a set of node pairs $\mathcal{P} = \{ (v_i, v_j) \}$ sampled from a graph $\mathcal{G}$, we construct another auxiliary graph $\mathcal{\tilde G}$, from which a feasible CD result of $\mathcal{G}$ can be extracted.
Each connected component in $\mathcal{\tilde G}$ corresponds to a community in $\mathcal{G}$, with the number of components as the estimated $K$.

\textit{\textbf{A novel learning paradigm}}.
Inspired by recent advances in graph pre-training techniques, PRoCD follows a novel \textit{pre-training \& refinement} paradigm with three phases as in Fig.~\ref{Fig:Overview} (c).
Based on the assumption that one has enough time to prepare a well-trained model in an offline way (i.e., \textit{offline pre-training}), we first pre-train PRoCD on small synthetic graphs with various topology properties (e.g., degree distributions) and high-quality community ground-truth.
We then generalize PRoCD (with frozen parameters) to large real graphs $\{\mathcal{G}'\}$ (i.e., \textit{online generalization}) and derive corresponding CD results $\{ \mathcal{\bar C'} \}$ via only one feed-forward propagation (FFP) of the model (i.e., inductive inference of DGL).
In some pre-training techniques \cite{wu2021self,sun2023graph}, \textit{online generalization} only provides an initialization of model parameters, which is further fine-tuned w.r.t. different tasks.
Analogous to this motivation, we treat $\mathcal{\bar C'}$ as the initialization of an efficient CD method (e.g., \textit{InfoMap} \cite{rosvall2008maps}) and adopt its output $\mathcal{C}'$ as the final CD result, which refines $\mathcal{\bar C'}$ (i.e., \textit{online refinement}).
In particular, the \textit{online generalization} and \textit{refinement} may benefit from the powerful transfer ability of inductive inference regarding quality while ensuring high inference efficiency, since there is no time-consuming model optimization.

Note that our \textit{pre-training \& refinement} paradigm is different from existing graph pre-training techniques \cite{wu2021self,sun2023graph}. We argue that they may suffer from the following issues about CD.
Besides pre-training, these methods may require another optimization procedure for the fine-tuning or prompt-tuning of specific tasks on $\{ \mathcal{G}'\}$, which is time-consuming and thus cannot help achieve high efficiency on $\{ \mathcal{G}'\}$.
To the best of our knowledge, related studies \cite{liu2023graphprompt,cao2023pre,sun2023all} merely focus on supervised tasks (e.g., node classification) and do not provide unsupervised tuning objectives for CD.

\textit{\textbf{A better trade-off between quality and efficiency}}.
Experiments on datasets with various scales demonstrate that PRoCD can ensure higher inference efficiency in CD without significant quality degradation, compared with running a refinement method from scratch. In some cases, PRoCD can even achieve improvement in both aspects.
Therefore, we believe that this study provides a possible way to obtain a better trade-off between the two aspects.

\begin{table*}[]\scriptsize
\caption{Summary of some representative CD methods.}\label{Tab:Meth}
\begin{tabular}{l|lllllllll|lllllll}
\hline
\textbf{Methods} & \multicolumn{1}{l|}{\textit{FastCom}} & \multicolumn{1}{l|}{\textit{GraClus}} & \multicolumn{1}{l|}{\textit{MC-SBM}} & \multicolumn{1}{l|}{\textit{Par-SBM}} & \multicolumn{1}{l|}{\textit{GMod}} & \multicolumn{1}{l|}{\textit{LPA}} & \multicolumn{1}{l|}{\textit{InfoMap}} & \multicolumn{1}{l|}{\textit{Louvain}} & \textit{Locale} & \multicolumn{1}{l|}{\textit{ClusNet}} & \multicolumn{1}{l|}{\textit{LGNN}} & \multicolumn{1}{l|}{\textit{GAP}} &  \multicolumn{1}{l|}{\textit{DMoN}} & \multicolumn{1}{l|}{\textit{DNR}} & \multicolumn{1}{l|}{\textit{DCRN}} & \textit{ICD} \\ \hline
\textbf{References} & \multicolumn{1}{l|}{\cite{clauset2004finding}} & \multicolumn{1}{l|}{\cite{dhillon2007weighted}} & \multicolumn{1}{l|}{\cite{peixoto2014efficient}} & \multicolumn{1}{l|}{\cite{peng2015scalable}} & \multicolumn{1}{l|}{\cite{clauset2004finding}} & \multicolumn{1}{l|}{\cite{raghavan2007near}} & \multicolumn{1}{l|}{\cite{rosvall2008maps}} & \multicolumn{1}{l|}{\cite{blondel2008fast}} & \cite{wang2020community} & \multicolumn{1}{l|}{\cite{wilder2019end}} & \multicolumn{1}{l|}{\cite{chen2019supervised}} & \multicolumn{1}{l|}{\cite{nazi2019gap}} & \multicolumn{1}{l|}{\cite{tsitsulin2023graph}} & \multicolumn{1}{l|}{\cite{yang2016modularity}} & \multicolumn{1}{l|}{\cite{liu2022deep}} & \cite{qin2023towards} \\ \hline
\textbf{Focus} & \multicolumn{9}{c|}{High Efficiency} & \multicolumn{6}{c|}{High Quality} & Q\&E \\ \hline
\textbf{Techniques} & \multicolumn{9}{c|}{Heuristic strategies or fast approximation w.r.t. relaxed CD objectives} & \multicolumn{4}{c|}{\textit{Task-dependent} DGL} & \multicolumn{3}{c}{\textit{Task-independent} DGL} \\ \hline
\textbf{Inference} & \multicolumn{9}{c|}{Running the algorithm from scratch for each graph} & \multicolumn{3}{c|}{\textit{Transductive} \& \textit{Inductive}} & \multicolumn{3}{c|}{\textit{Transductive}} & \textit{Inductive} \\ \hline
$K$-\textbf{agnostic} & \multicolumn{1}{l|}{\usym{2718}} & \multicolumn{1}{l|}{\usym{2718}} & \multicolumn{1}{l|}{\usym{2714}} & \multicolumn{1}{l|}{\usym{2714}} & \multicolumn{1}{l|}{\usym{2714}} & \multicolumn{1}{l|}{\usym{2714}} & \multicolumn{1}{l|}{\usym{2714}} & \multicolumn{1}{l|}{\usym{2714}} & \usym{2714} & \multicolumn{1}{l|}{\usym{2718}} & \multicolumn{1}{l|}{\usym{2718}} & \multicolumn{1}{l|}{\usym{2718}} & \multicolumn{1}{l|}{\usym{2718}} & \multicolumn{1}{l|}{\usym{2718}} & \multicolumn{1}{l|}{\usym{2718}} & \usym{2718} \\ \hline
\end{tabular}
\end{table*}

\section{Related Work}\label{Sec:Rel}

In the past few decades, many CD methods have been proposed based on different problem statements, hypotheses, and techniques \cite{xing2022comprehensive}. Table~\ref{Tab:Meth} summarizes some representative approaches according to their original designs. Most of them focus on either high efficiency or quality.
Conventional efficient CD methods use heuristic strategies or fast approximation w.r.t. some relaxed CD objectives to derive feasible CD results, including the (\romannumeral1) greedy modularity maximization in \textit{FastCom} \cite{clauset2004finding} and \textit{Louvain} \cite{blondel2008fast}, (\romannumeral2) semi-definite relaxation of modularity in \textit{Locale} \cite{wang2020community}, (\romannumeral3) label propagation heuristic in \textit{LPA} \cite{raghavan2007near}, as well as (\romannumeral4) Monte Carlo approximation of stochastic block model (SBM) in \textit{MC-SBM} \cite{peixoto2014efficient} and \textit{Par-SBM} \cite{peng2015scalable}.

Recent studies have also demonstrated the powerful potential of DGL to ensure high quality of CD, following the architectures in Fig.~\ref{Fig:Overview} (a) and (b).
However, some methods (e.g., \textit{DNR} \cite{yang2016modularity}, \textit{DCRN} \cite{liu2022deep}, and \textit{DMoN} \cite{tsitsulin2023graph}) only considered the inefficient \textit{transductive} inference, with time-consuming model optimization in the inference of CD on each single graph.
Other approaches (e.g., \textit{ClusNet} \cite{wilder2019end}, \textit{LGNN} \cite{chen2019supervised}, \textit{GAP} \cite{nazi2019gap}, and \textit{ICD} \cite{qin2023towards}) considered \textit{inductive} inference across graphs.
Results of \textit{ICD} validated that the \textit{inductive} inference can help achieve a better trade-off between quality and efficiency in \textit{online generalization} (i.e., directly generalizing a model pre-trained on historical graphs $\{ \mathcal{G}_t \}$ to new graphs $\{ \mathcal{G}' \}$).
Nevertheless, the quality of \textit{online generalization} may be affected by the possible distribution shift between $\{ \mathcal{G}_t \}$ and $\{ \mathcal{G}' \}$.
In contrast, \textit{online generalization} is used to construct the initialization of \textit{online refinement} in our PRoCD method, which can ensure better quality on $\{ \mathcal{G}' \}$.
Moreover, most \textit{task-dependent} DGL methods (e.g., \textit{ClusNet}, \textit{LGNN}, \textit{GAP}, and \textit{DMoN}) are inapplicable to $K$-agnostic CD, since they usually contain an output module related to $K$. Although \textit{task-independent} approaches (e.g., \textit{DNR}, \textit{DCRN}, and \textit{ICD}) do not contain such a module, their original designs still rely on a pre-set $K$ (e.g., $K$Means as the downstream clustering algorithm).

Different from the aforementioned methods, our PRoCD method can ensure higher efficiency in $K$-agnostic CD without significant quality degradation via a novel model architecture following the \textit{pre-training \& refinement} paradigm, as highlighted in Fig.~\ref{Fig:Overview} (c).

As reviewed in \cite{wu2021self,sun2023graph}, existing graph pre-training techniques usually follow the paradigms of (\romannumeral1) \textit{pre-training \& fine-tuning} (e.g., \textit{GCC} \cite{qiu2020gcc}, \textit{L2P-GNN} \cite{lu2021learning}, and \textit{W2P-GNN} \cite{cao2023pre}) as well as (\romannumeral2) \textit{pre-training \& prompting} (e.g., \textit{GPPT} \cite{sun2022gppt}, \textit{GraphPrompt} \cite{liu2023graphprompt}, and \textit{ProG} \cite{sun2023all}).
In addition to \textit{offline pre-training}, these methods rely on another optimization procedure for the fine-tuning or prompt-tuning of specific inference tasks, which is usually time-consuming. Moreover, most of them merely focus on supervised tasks (e.g., node classification, link prediction, and graph classification) and do not provide unsupervised tuning objectives for CD.
Therefore, one cannot directly apply these pre-training methods to ensure the high efficiency in CD without quality degradation.

\section{Problem Statements \& Preliminaries}\label{Sec:Prob}
In this study, we consider the disjoint CD on undirected graphs. A graph can be represented as $\mathcal{G} = (\mathcal{V}, \mathcal{E})$, with $\mathcal{V} = \{ v_1, \cdots, v_N\}$ and $\mathcal{E} = \{ (v_i, v_j)| v_i, v_j \in \mathcal{V})\}$ as the sets of nodes and edges. The topology of $\mathcal{G}$ can be described by an adjacency matrix ${\bf{A}} \in \{ 0, 1\} ^ {N \times N}$, where ${\bf{A}}_{ij} = {\bf{A}}_{ji} = 1$ if $(v_i, v_j) \in \mathcal{E}$ and ${\bf{A}}_{ij} = {\bf{A}}_{ji} = 0$ otherwise. Since CD is originally defined only based on graph topology \cite{fortunato202220}, we assume that graph attributes are unavailable.

\textbf{Community Detection} (CD). Given a graph $\mathcal{G}$, CD aims to partition the node set $\mathcal{V}$ into $K$ subsets (defined as communities) ${\mathcal{C}} = \{ \mathcal{C}_1, \cdots, \mathcal{C}_K\}$ (${C_r} \cap {C_s} = \emptyset $ for $\forall r \ne s$) s.t. (\romannumeral1) the linkage within each community is dense but (\romannumeral2) that between communities is relatively loose.
We define that a CD task is $K$-agnostic if the number of communities $K$ is unknown for a given graph $\mathcal{G}$, where one should simultaneously determine $K$ and the corresponding CD result $\mathcal{C}$.
We consider $K$-agnostic CD in this study. 
Whereas, some existing methods \cite{dhillon2007weighted,wilder2019end,tsitsulin2023graph} assume that $K$ is given for each input ${\mathcal{G}}$ and thus cannot tackle such a challenging yet realistic setting.

\textbf{Modularity Maximization}.
Mathematically, CD can be formulated as the modularity maximization objective \cite{newman2006modularity}, which maximizes the difference between the exact graph topology (described by ${\bf{A}}$) and a randomly generated case.
Given $\mathcal{G}$ and $K$, it aims to derive a partition $\mathcal{C}$ that maximizes the following modularity metric:
\begin{equation}\label{Eq:Mod-Max}
    \mathop {\max }\limits_\mathcal{C} {\mathop{\rm Mod}\nolimits} (\mathcal{G}, K) := \frac{1}{{2M}}\sum\limits_{r = 1}^K {\sum\limits_{{v_i},{v_j} \in {\mathcal{C}_r}} {[{{\bf{A}}_{ij}} - \frac{{{{\deg} (v_i)}{{\deg} (v_j)}}}{{2M}}]} },
\end{equation}
where $\deg (v_i)$ is the degree of node $v_i$; $M$ is the number of edges. Objective (\ref{Eq:Mod-Max}) can be rewritten into the following matrix form:
\begin{equation}\label{Eq:Mod-Max-Mat}
    \mathop {\min }\limits_{\bf{H}}  - {\mathop{\rm tr}\nolimits} ({{\bf{H}}^T}{\bf{QH}}){\rm{~s.t.~}}{{\bf{H}}_{ir}} = \left\{ {\begin{array}{*{20}{l}}
    {1,~{v_i} \in {C_r}}\\
    {0,~{\rm{otherwise}}}
    \end{array}} \right.,
\end{equation}
where ${\bf{Q}} \in \mathbb{R}^{N \times N}$ is defined as the modularity matrix with ${{\bf{Q}}_{ij}} := [{{\bf{A}}_{ij}} - {{\deg} (v_i)}{{\deg} (v_j)} /(2M)]$; ${\bf{H}} \in \{0, 1\}^{N \times K}$ indicates the community membership of $\mathcal{C}$.
Our method does not directly solve the aforementioned problem. Instead, we try to extract informative community-preserving features from ${\bf{Q}}$.

\textbf{\textit{Pre-training} \& \textit{Refinement} Paradigm}.
To achieve a better trade-off between the quality and efficiency, we propose a \textit{pre-training \& refinement} paradigm based on the inductive inference of DGL.
We represent a DGL model as $\mathcal{C} = f(\mathcal{G}; \Theta )$, which derives the CD result $\mathcal{C}$ given a graph $\mathcal{G}$, with $\Theta$ as the set of trainable model parameters.
As in Fig.~\ref{Fig:Overview} (c), the proposed paradigm includes (\romannumeral1) \textit{offline pre-training}, (\romannumeral2) \textit{online generalization}, and (\romannumeral3) \textit{online refinement}.

In \textit{offline pre-training}, we first generate a set of synthetic graphs $\mathcal{T} = \{ \mathcal{G}_1, \cdots, \mathcal{G}_T \}$ via a generator (e.g., SBM \cite{karrer2011stochastic}). The generation of each graph $\mathcal{G}_t \in \mathcal{T}$ includes the topology $(\mathcal{V}_t, \mathcal{E}_t)$ and community assignment ground-truth $\mathcal{C}^{(t)}$. We then pre-train $f$ on $\mathcal{T}$ (i.e., updating $\Theta$) based on an objective regarding $\{ (\mathcal{V}_t, \mathcal{E}_t) \}$ and $\{ \mathcal{C}^{(t)} \}$ in an \textit{offline} way.
After that, we generalize $f$ to a large real graph $\mathcal{G}'$ with frozen $\Theta$ (i.e., \textit{online generalization}), which can derive a feasible CD result $\mathcal{\bar C}'$ via only one FFP of $f$ (i.e., inductive inference). Analogous to the fine-tuning in existing pre-training techniques, we treat $\mathcal{\bar C}'$ as the initialization of a conventional efficient CD method (e.g., \textit{InfoMap} \cite{rosvall2008maps}) and adopt its output $\mathcal{C}'$ as the final CD result, which further refines $\mathcal{\bar C}'$ (i.e., \textit{online refinement}).

\textbf{Evaluation Protocol}.
In real applications, it is usually assumed that \textit{one has enough time to prepare a well-trained model in an offline way} (e.g., pre-training of LLMs).
After deploying $f$, one may achieve high efficiency in the online inference on graphs $\{ \mathcal{G}' \}$ (e.g., via one FFP of $f$ without any model optimization).
In contrast, conventional methods cannot benefit from pre-training but have to run their algorithms on $\{ \mathcal{G}' \}$ from scratch, due to the inapplicability to inductive inference.
Our evaluation focuses on the \textit{online inference} of CD on $\{ \mathcal{G}' \}$ (e.g., \textit{online generalization} and \textit{refinement} of PRoCD), which is analogous to the applications of foundation models.
For instance, users benefit from the powerful online inference of LLMs (e.g., generating high-quality answers in few seconds) but do not need to train them using a great amount of resources.

\section{Methodology}\label{Sec:Meth}
We propose PRoCD following a novel \textit{pre-training \& refinement} paradigm as highlighted in Fig.~\ref{Fig:Overview} (c).

\subsection{Model Architecture}

\begin{figure}[t]
  \centering
  \includegraphics[width=\linewidth, trim=23 32 18 18,clip]{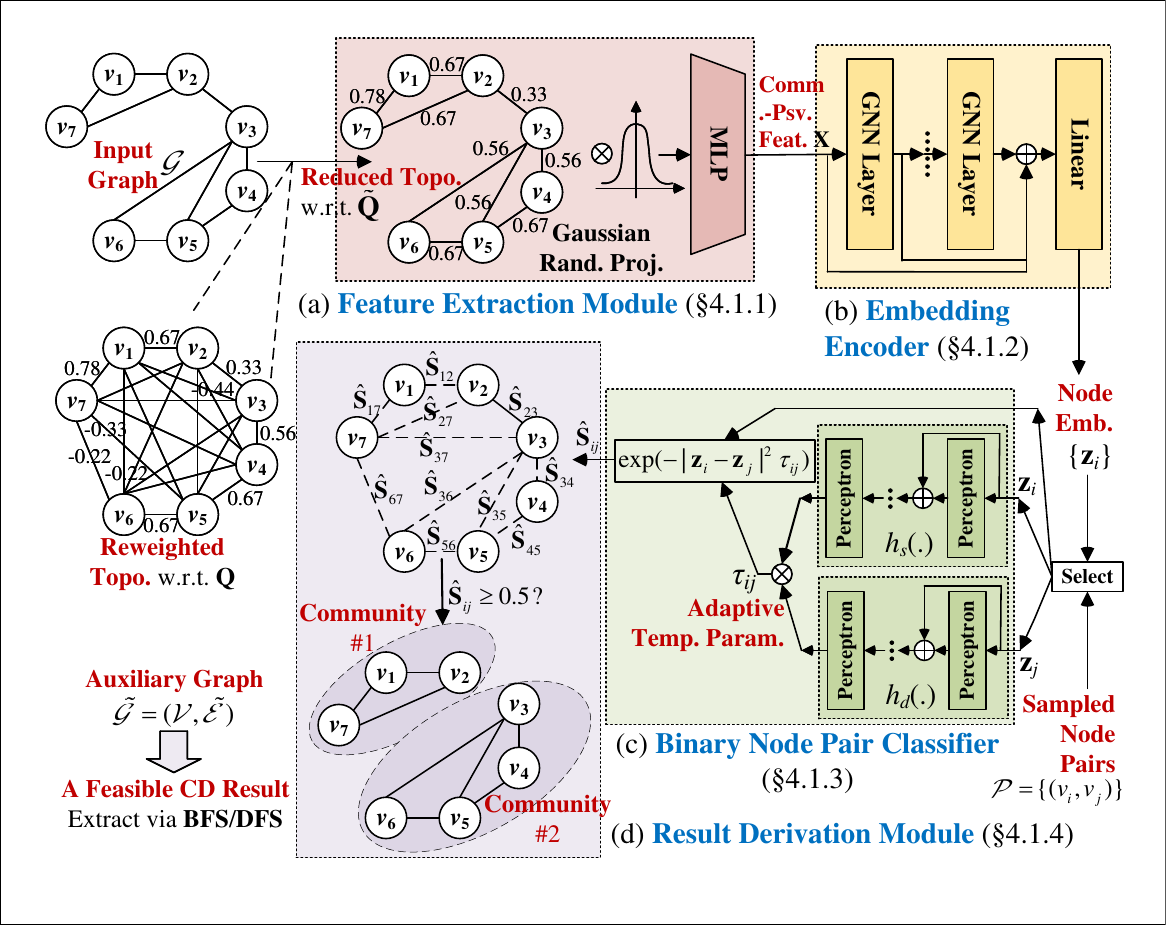}
  \caption{Overview of the model architecture of PRoCD.}\label{Fig:Mdl-Arch}
\end{figure}

To enable PRoCD to derive feasible results for $K$-agnostic CD in an end-to-end architecture, we reformulate CD as the binary node pair classification.
For each graph $\mathcal{G} = (\mathcal{V}, \mathcal{E})$, we introduce an auxiliary variable ${\bf{S}} \in \{ 0, 1\} ^{N \times N}$ ($N := |\mathcal{V}|$), where ${\bf{S}}_{ij} = {\bf{S}}_{ji} = 1$ if nodes $v_i$ and $v_j$ are partitioned into the same community and ${\bf{S}}_{ij} = {\bf{S}}_{ji} = 0$ otherwise.
Given a set of node pairs $\mathcal{P} = \{ (v_i, v_j) \}$, we also rearrange corresponding elements in ${\bf{S}}$ as a vector ${\bf{y}} \in \{ 0, 1\}^{|\mathcal{P}|}$, where ${\bf{y}}_l = {\bf{S}}_{ij} = {\bf{S}}_{ji}$ w.r.t. each node pair $p_l = (v_i, v_j) \in \mathcal{P}$.
Fig.~\ref{Fig:Mdl-Arch} provides an overview of the model architecture with a running example.
Given a graph $\mathcal{G}$, the model samples a set of node pairs $\mathcal{P}$ and derives the estimated value of ${\bf{y}}_l = {\bf{S}}_{ij}$ w.r.t. each $p_l = (v_i, v_j) \in \mathcal{P}$, from which a feasible CD result $\mathcal{C}$ can be extracted.

\subsubsection{\textbf{Feature Extraction Module}}
As shown in Fig.~\ref{Fig:Mdl-Arch} (a), we first extract community-preserving features, arranged as ${\bf{X}} \in \mathbb{R}^{N \times d}$ ($d \ll N$), for an input graph $\mathcal{G}$ from the modularity maximization objective (\ref{Eq:Mod-Max-Mat}).
The modularity matrix ${\bf{Q}}$ in (\ref{Eq:Mod-Max-Mat}) is a primary component regarding graph topology. It can be considered as a reweighting of original topology, where nodes $(v_i, v_j)$ with similar neighbor-induced features $({\bf{Q}}_{i,:}, {\bf{Q}}_{j,:})$ are more likely to belong to a common community. To minimize the objective in (\ref{Eq:Mod-Max-Mat}), which is equivalent to maximizing the modularity metric in (\ref{Eq:Mod-Max}), ${\bf{Q}}_{ij}$ with a large value indicates that $(v_i, v_j)$ are more likely to be partitioned into the same community (e.g., ${\bf{Q}}_{17} = {\bf{Q}}_{71} = 0.78$ in Fig.~\ref{Fig:Mdl-Arch}).
Therefore, we believe that ${\bf{Q}}$ encodes key characteristics regarding community structures.

Note that ${\bf{Q}} \in \mathbb{R}^{N \times N}$ is usually dense. To utilize the sparsity of topology, we reduce ${\bf{Q}}$ to a sparse matrix ${\bf{\tilde Q}} \in \mathbb{R}^{N \times N}$, where ${\bf{\tilde Q}}_{ij} = {\bf{Q}}_{ij}$ if $(v_i, v_j) \in \mathcal{E}$ and ${\bf{\tilde Q}}_{ij} = 0$ otherwise. Although the reduction may lose some information, it enables the model to be scaled up to large sparse graphs without constructing an $N \times N$ dense matrix. Our experiments demonstrate that PRoCD can still derive high-quality CD results using ${\bf{\tilde Q}}$. Given ${\bf{\tilde Q}}$, we derive ${\bf{X}}$ via
\begin{equation}\label{Eq:Feat-Ext}
    {\bf{X}} = {\mathop{\rm MLP}\nolimits} ({\bf{\tilde Q \Omega }}), {\rm{~with~}}{\bf{\Omega }} \in {\mathbb{R}^{N \times d}} \sim \mathcal{N} (0, 1/d).
\end{equation}
Concretely, the Gaussian random projection \cite{arriaga2006algorithmic}, an efficient dimension reduction technique that can preserve the relative distances between input features with rigorous theoretical guarantees, is first applied to ${\bf{\tilde Q}}$. We then incorporate non-linearity into the reduced features using an MLP.

\subsubsection{\textbf{Embedding Encoder}}
Given the extracted features ${\bf{X}}$, we then derive low-dimensional node embeddings $\{ {\bf{z}}_i \in \mathbb{R}^d \}$ using a multi-layer GNN with skip connections, as shown in Fig.~\ref{Fig:Mdl-Arch} (b). Here, we adopt GCN \cite{kipf2016semi} as an example building block. One can easily extend the model to include other advanced GNNs. Let ${\bf{Z}}^{[s-1]}$ and ${\bf{Z}}^{[s]}$ be the input and output of the $s$-th GNN layer, with ${\bf{Z}}^{[0]} = {\bf{X}}$. The multi-layer GNN can be described as
\begin{equation}\label{Eq:Emb-Enc}
    \begin{array}{l}
    {\bf{Z}} = {\mathop{\rm LN}\nolimits} ({\mathop{\rm Linear}\nolimits} ({\bf{\tilde Z}})),~{\bf{\tilde Z}} = \sum\nolimits_s {{{\bf{Z}}^{[s]}}} ,\\
    {{\bf{Z}}^{[s]}} = {\mathop{\rm LN}\nolimits} (\tanh ({{{\bf{\hat D}}}^{ - 0.5}}{\bf{\hat A}}{{{\bf{\hat D}}}^{ - 0.5}}{{\bf{Z}}^{[s - 1]}}{\bf{W}}{}^{[s]})),
\end{array}
\end{equation}
where ${\bf{\hat A}} := {\bf{A}} + {\bf{I}}_N$ is the adjacency matrix containing self-edges; ${\bf {\hat D}}$ is the degree diagonal matrix w.r.t. ${\bf{\hat A}}$; ${\bf{W}}^{[s]} \in \mathbb{R}^{d \times d}$ is a trainable weight matrix; ${\mathop{\rm LN}\nolimits} ({\bf{U}})$ denotes the row-wise $l_2$ normalization on a matrix ${\bf{U}}$ (i.e., ${{\bf{U}}_{i,:}} \leftarrow {{\bf{U}}_{i,:}}/|{{\bf{U}}_{i,:}}{|_2}$); 
${\bf{Z}} \in \mathbb{R}^{N \times d}$ is the matrix form of the derived embeddings, with ${\bf{Z}}_{i, :} = {\bf{z}}_i \in \mathbb{R}^d$ as the embedding of node $v_i$.
In each GNN layer, ${\bf{Z}}_{i,:}^{[s]}$ is an intermediate representation of node $v_i$, which is the non-linear aggregations (i.e., weighted mean) of features w.r.t. $\{ v_i \} \cup {\mathop{\rm Nei}\nolimits} ({v_i})$, with ${\mathop{\rm Nei}\nolimits} ({v_i})$ as the set of neighbors of $v_i$. Since this aggregation operation forces nodes $(v_i, v_j)$ with similar neighbors $({\mathop{\rm Nei}\nolimits} ({v_i}), {\mathop{\rm Nei}\nolimits} ({v_j}))$ (i.e., dense local topology) to have similar representations $({\bf{Z}}_{i,:}^{[s]}, {\bf{Z}}_{j,:}^{[s]})$, the multi-layer GNN can enhance the ability of PRoCD to derive community-preserving embeddings. To obtain ${\bf{Z}}$, we first sum up the intermediate representations $\{ {\bf{Z}}^{[s]} \}$ of all the layers and apply a linear mapping to the summed representation ${\bf{\tilde Z}}$. Furthermore, the row-wise $l_2$ normalization is applied. It forces $\{ {\bf{z}}_i \}$ to be distributed in a unit sphere, with $|{{\bf{z}}_i} - {{\bf{z}}_j}{|^2} = 2 - 2{{\bf{z}}_i}{\bf{z}}_j^T$.

\subsubsection{\textbf{Binary Node Pair Classifier}}
As highlighted in Fig.~\ref{Fig:Mdl-Arch} (c), given a pair of nodes $(v_i, v_j)$ sampled from a graph $\mathcal{G}$, we use a binary classifier to estimate ${\bf{S}}_{ij}$ based on embeddings $({\bf{z}}_i, {\bf{z}}_j)$ and determine whether $(v_i, v_j)$ are partitioned into the same community.
A widely adopted design of binary classifier is as follow
\begin{equation}\label{Eq:Naive-Bin-Clf}
    {{{\bf{\hat S}}}_{ij}} = {\mathop{\rm sigmoid}\nolimits} ({{\bf{z}}_i}{{\bf{z}}_j^T}/\tau ),
\end{equation}
with $\tau$ as a pre-set temperature parameter. Instead of directly using (\ref{Eq:Naive-Bin-Clf}), we introduce a novel binary classifier with \textit{pair-wise adaptive temperature parameters} $\{ \tau_{ij} \}$, which can derive a more accurate estimation of ${\bf{S}}$ to derive high-quality CD results as demonstrated in our experiments. Our binary classifier can be described as
\begin{equation}\label{Eq:Bin-Clf}
    \begin{array}{l}
    {{{\bf{\hat S}}}_{ij}} = \exp ( - |{{\bf{z}}_i} - {{\bf{z}}_j}{|^2}{\tau _{ij}}) = \exp (2{\tau _{ij}}({{\bf{z}}_i}{\bf{z}}_j^T - 1)),\\
    {\rm{with~}}{\tau _{ij}} = {h_s}({{\bf{z}}_i}){h_d}{({{\bf{z}}_j})^T},
    \end{array}
\end{equation}
where $h_s$ and $h_d$ are MLPs with the same configuration.
In (\ref{Eq:Bin-Clf}), ${\bf{S}}_{ij}$ is estimated based on the distance between $({\bf{z}}_i, {\bf{z}}_j)$ and a pair-wise temperature parameter $\tau_{ij}$. Different node pairs $\{ (v_i, v_j) \}$ may have different parameters $\{ \tau_{ij} \}$ determined by embeddings $\{ ({\bf{z}}_i, {\bf{z}}_j) \}$.

\subsubsection{\textbf{Result Derivation Module}}

\begin{algorithm}[t]\footnotesize
\caption{Deriving a Feasible CD Result}
\label{Alg:Post-Prep}
\LinesNumbered
\KwIn{input graph $\mathcal{G} = (\mathcal{V}, \mathcal{E})$; number of sampled node pairs $n_S$}
\KwOut{a feasible CD result $\mathcal{C}$ w.r.t. $\mathcal{G}$}
$\mathcal{P} \leftarrow \mathcal{E}$//Initialize the set of node pairs $\mathcal{P}$ using edge set $\mathcal{E}$\\
\For{sample\_count \textbf{from} $1$ \textbf{to} $n_S$}
{
    randomly sample a node pair $p = (v_i, v_j)$\\
    add the sampled $p = (v_i, v_j)$ to $\mathcal{P}$\\
}
derive $\bf{\hat y}$ w.r.t. $\mathcal{P}$ via one FFP of the model\\
$\mathcal{\tilde E} \leftarrow \emptyset$//Initialize edge set of auxiliary graph $\mathcal{\tilde G} = (\mathcal{V}, \mathcal{\tilde E})$\\
\For{{\bf{each}} {\rm{node pair}} $p_l = (v_i, v_j) \in \mathcal{P}$}
{
    \If{${\bf{\hat y}}_l > 0.5$}
    {
        add $p_l = (v_i, v_j)$ to $\mathcal{\tilde E}$
    }
}
extract connected components of $\mathcal{\tilde G}$ via DFS/BFS on $\mathcal{\tilde E}$\\
treat each extracted component as a community to form $\mathcal{C}$
\end{algorithm}

As illustrated in Fig.~\ref{Fig:Mdl-Arch} (d), we develop a result derivation module, which outputs a feasible CD result $\mathcal{C}$ based on a set of node pairs sampled from the input graph $\mathcal{G}$. Algorithm~\ref{Alg:Post-Prep} summarizes the procedure of this module.

In lines 1-4, we construct a set of node pairs $\mathcal{P} = \{ (v_i, v_j) \}$ (e.g., dotted lines in Fig.~\ref{Fig:Mdl-Arch} (d)), which includes (\romannumeral1) all the edges of input graph $\mathcal{G}$ (e.g., $(v_1, v_7)$ and $(v_3, v_4)$ in Fig.~\ref{Fig:Mdl-Arch} (d)) and (\romannumeral2) $n_S$ randomly sampled node pairs (e.g., $(v_6, v_7)$ and $(v_3, v_7)$ in Fig.~\ref{Fig:Mdl-Arch} (d)).

In line 5, we derive the estimated values $\{ {\bf{\hat S}}_{ij} \}$ w.r.t. node pairs in $\mathcal{P}$ (via one FFP of the model) and rearrange them as a vector ${\bf{\hat y}} \in \mathbb{R}^{|\mathcal{P}|}$.
In particular, ${\bf{\hat y}}_l = {\bf{\hat S}}_{ij}$ represents the probability that a pair of nodes $p_l = (v_i, v_j)$ are partitioned into a common community.

In lines 6-9, we further construct an auxiliary graph ${\mathcal{\tilde G}} = (\mathcal{V}, \mathcal{\tilde E})$ based on ${\bf{\hat y}}$ and $\mathcal{P}$. $\mathcal{\tilde G}$ has the same node set $\mathcal{V}$ as the input graph $\mathcal{G}$ but a different edge set $\mathcal{\tilde E}$. For each $p_l = (v_i, v_j) \in \mathcal{P}$, we add $(v_i, v_j)$ to $\mathcal{\tilde E}$ (i.e., preserving this node pair as an edge in $\mathcal{\tilde G}$) if ${\bf{\hat y}}_l > 0.5$.
$\mathcal{\tilde G}$ may contain multiple connected components (e.g., the two components in Fig.~\ref{Fig:Mdl-Arch} (d)). All the edges $\{ e_l = (v_i, v_j) \}$ within a component are with high values of $\{ {\bf{\hat y}}_l = {\bf{\hat S}}_{ij} \}$, indicating that the associated nodes are more likely to belong to the same community.

In lines 10-11, we derive a feasible CD result $\mathcal{C}$ w.r.t. $\mathcal{G}$ by extracting all the connected components of ${\mathcal{\tilde G}}$ via the depth-first search (DFS) or breadth-first search (BFS).
Each connected component of ${\mathcal{\tilde G}}$ corresponds to a unique community in ${\mathcal{G}}$ (e.g., the two communities in Fig.~\ref{Fig:Mdl-Arch} (d)), with the number of components as the estimated number of communities $K$. Therefore, the aforementioned designs enable our PRoCD method to tackle $K$-agnostic CD.

\subsection{Offline Pre-Training}
We conducted the \textit{offline pre-training} of PRoCD on a set of synthetic graphs $\mathcal{T}  = \{ {\mathcal{G}}_1, \cdots, {\mathcal{G}}_T \}$.
One significant advantage of using synthetic pre-training data is that we can simulate various properties of graph topology with different levels of noise and high-quality ground-truth by adjusting parameters of the synthetic generator.

In particular, we consider a challenging setting of pre-training PRoCD on small synthetic graphs (e.g., with $N \approx 5 \times 10^3$ nodes) but generalize it to large real graphs (e.g., $N > 10^6$).
One reason of using small pre-training graphs is that it enables PRoCD to construct dense $N \times N$ matrices in pre-training objectives (e.g., ${\bf{\hat S}} \in \mathbb{R}^{N \times N}$ for binary classification) and thus fully explore the topology and community ground-truth of pre-training data.
In contrast, constructing dense $N \times N$ matrices for large graphs may be intractable.
Although there may be distribution shifts between the (\romannumeral1) small pre-training graphs $\{ \mathcal{G}_t \}$ and (\romannumeral2) large graphs $\{ \mathcal{G}' \}$ to be partitioned, our experiments demonstrate that \textit{offline pre-training} on $\{ \mathcal{G}_t \}$ is essential for PRoCD to derive feasible CD results $\{ \mathcal{\bar C}' \}$ for $\{ \mathcal{G}' \}$. By using $\{ \mathcal{\bar C}' \}$ as the initialization, their quality can be further refined via an efficient CD method.

\subsubsection{\textbf{Generation of Pre-Training Data}}\label{Sec:PTN-Data-Gen}
As a demonstration, we adopt the degree-corrected SBM (DC-SBM) \cite{karrer2011stochastic} implemented by \texttt{graph-tool}\footnote{https://graph-tool.skewed.de/} to generate synthetic pre-training graphs.
Besides topology, the generator also produces community assignments w.r.t. a specified level of noise, which can be used as the high-quality ground-truth of CD.
Our experiments also validate that the synthetic ground-truth can help PRoCD derive a good initialization for \textit{online refinement}. Whereas, community ground-truth is usually unavailable for existing methods \cite{wilder2019end,tsitsulin2023graph} (pre-)trained on real graphs.
Instead of using fixed generator parameters, we let them follow certain probability distributions to simulate various topology properties (e.g., distributions of node degrees and community sizes) and noise levels of community structures.
Namely, different graphs may be generated based on different parameter settings sampled from certain distributions.
Due to space limits, we leave details about the parameter setting and generation algorithm in Appendix~\ref{App-Syn-Alg}.

\subsubsection{\textbf{Pre-Training Objectives \& Algorithms}}
For each synthetic graph $\mathcal{G}_t$, suppose there are $N_t$ nodes partitioned into $K_t$ communities. We use ${\bf{M}}^{(t)}$ to denote a vector/matrix associated with ${\mathcal{G}}_t$ (e.g., ${\bf{\hat y}}^{(t)}$ and $\bf{\hat S}^{(t)}$).
Given a graph $\mathcal{G}_t$, some previous studies \cite{wilder2019end,tsitsulin2023graph} adopt a relaxed version of the modularity maximization objective (\ref{Eq:Mod-Max-Mat}) for model optimization and demonstrate its potential to capture community structures.
This relaxed objective allows ${\bf{H}}^{(t)} \in \{ 0, 1\}^{N_t \times K_t}$ in (\ref{Eq:Mod-Max-Mat}), which describes hard community assignments, to be continuous values (e.g., ${\bf{H}}^{(t)} \in \mathbb{R}_+^{N_t \times K_t}$ derived via an MLP).
Note that we reformulate CD as the binary node pair classification with ${\bf{S}}^{(t)} \in \{ 0, 1\}^{N_t \times N_t}$. ${\bf{\hat S}}^{(t)} \in \mathbb{R}_+^{N_t \times N_t}$ derived from the binary classifier can be considered as the relaxation of ${\bf{S}}^{(t)}$. Based on ${\bf{\hat S}}^{(t)}$, we rewrite the modularity maximization objective (\ref{Eq:Mod-Max}) to the following relaxed form for each graph $\mathcal{G}_t$: 
\begin{equation}\label{Eq:Mod-Obj}
    {\mathcal{L}_{\rm MOD}} (\mathcal{G}_t) :=  - [\frac{1}{M}\sum\limits_{{e_l} \in {\mathcal{E}_t}} {{\bf{\hat y}}_l^{(t)}}  - \frac{\lambda}{{4{M^2}}}\sum\limits_{ij} {{{({{\bf{D}}^{(t)}}{{\bf{\hat S}}^{(t)}}{{\bf{D}}^{(t)}})}_{ij}}} ],
\end{equation}
where ${\bf{\hat y}}^{(t)}$ is the rearrangement of elements in ${\bf{\hat S}}^{(t)}$ w.r.t. the edge set $\mathcal{E}_t$; ${\bf{D}}^{(t)}$ is the degree diagonal matrix of $\mathcal{G}_t$; $\lambda > 0$ is the resolution parameter of modularity maximization \cite{reichardt2006statistical}, with $\lambda <1$ ($>1$) favoring the partition of large (small) communities. 

In addition, we utilize the synthetic community ground-truth $\mathcal{C}_t$ of each graph $\mathcal{G}_t$ to enhance the ability of PRoCD to capture community structures.
Note that different graphs $\{ \mathcal{G}_t \}$ may have different numbers of communities $\{ K_t \}$. However, some DGL-based CD methods \cite{wilder2019end,chen2019supervised,tsitsulin2023graph} are only designed for graphs with a fixed $K$.
Community labels are also permutation-invariant. For instance, $(c_1, c_2, c_3, c_4, c_5) = (1, 1, 1, 2, 2)$ and $(c_1, c_2, c_3, c_4, c_5) = (2, 2, 2, 1, 1)$ represent the same community assignment, with $c_i$ as the label assignment of node $v_i$. Hence, the standard multi-class cross entropy objective cannot be directly applied to integrate community ground-truth. PRoCD handles these issues by reformulating CD as the binary node pair classification with auxiliary variables $\{ {\bf{S}}^{(t)} \}$, where the dimensionality and values of $\{ {\bf{S}}^{(t)} \}$ are unrelated to $\{ K_t \}$ and permutations of community labels.
Given a graph ${\mathcal{G}}_t$, one can construct ${\bf{S}}^{(t)}$ based on ground-truth ${\mathcal{C}}^{(t)}$ and derive ${\bf{\hat S}}^{(t)}$ via one FFP of the model. We introduce the following binary cross entropy objective that covers all the $N_t \times N_t$ node pairs:
\begin{equation}\label{Eq:BCE-Obj}
    {\mathcal{L}_{\rm BCE}} (\mathcal{G}_t) :=  - \sum\limits_{{v_i},{v_j} \in \mathcal{V}} {\left[ \begin{array}{l}
    {\bf{S}}_{ij}^{(t)}\log {\bf{\hat S}}_{ij}^{(t)} + \\
    (1 - {\bf{S}}_{ij}^{(t)})\log (1 - {\bf{\hat S}}_{ij}^{(t)})
\end{array} \right]}.
\end{equation}
Finally, we formulate the pre-training objective w.r.t. $\mathcal{G}_t$ as
\begin{equation}\label{Eq:PTN-Obj}
    {\mathcal{L}_{{\rm{PTN}}}}({\mathcal{G}_t}) := {\mathcal{L}_{{\rm{MOD}}}}({\mathcal{G}_t}) + \alpha {\mathcal{L}_{{\mathop{\rm BCE}\nolimits} }}({\mathcal{G}_t}),
\end{equation}
where $\alpha > 0$ is the hyper-parameter to balance ${\mathcal{L}_{{\rm{MOD}}}}$ and ${\mathcal{L}_{{\rm{BCE}}}}$. 
The \textit{offline pre-training} procedure is concluded in Algorithm~\ref{Alg:PTN}, where the Adam optimizer with learning rate $\eta$ is applied to update model parameters $\Theta$; the updated $\Theta$ after $n_P$ epochs are then saved for \textit{online generalization and refinement}.
Since only the embedding encoder and binary classifier contain model parameters to be learned, we do not need to apply Algorithm~\ref{Alg:Post-Prep} of the result derivation module to derive a feasible CD result in \textit{offline pre-training}.

\begin{algorithm}[t]\footnotesize
\caption{\textit{Offline Pre-training}}
\label{Alg:PTN}
\LinesNumbered
\KwIn{synthetic pre-training graphs $\mathcal{T} = \{ \mathcal{G}_1, \cdots, \mathcal{G}_T \}$ and their ground-truth; hyper-parameters $\{ \lambda, \alpha \}$; number of epochs $n_P$; learning rate $\eta$}
\KwOut{optimized model parameters $\Theta^*$}
initialize model parameters $\Theta$\\
\For{epoch \textbf{from} $1$ \textbf{to} $n_P$}
{
    \For{{\bf{each}} $\mathcal{G}_t = (\mathcal{V}_t, \mathcal{E}_t) \in \mathcal{T}$}
    {
        construct ${\bf{S}}^{(t)}$ w.r.t. ground-truth $\mathcal{C}^{(t)}$\\
        get input feature ${\bf{X}}^{(t)}$ w.r.t. $\mathcal{G}_t$ via (\ref{Eq:Feat-Ext})\\
        get estimated values ${\bf{\hat S}}^{(t)}$ w.r.t. $\mathcal{G}_t$ via one FFP\\
        arrange elements w.r.t. ${\mathcal{E}}_t$ in ${\bf{\hat S}}^{(t)}$ as ${\bf{\hat y}}^{(t)}$\\
        update model parameters $\Theta  \leftarrow {\mathop{\rm Opt}\nolimits} (\eta ,\Theta , \partial {\mathcal{L}_{{\rm{PTN}}}} (\mathcal{G}_t) / \partial \Theta )$
    }
}
save the optimized model parameters $\Theta^*$
\end{algorithm}

\subsection{Online Generalization \& Refinement}
After \textit{offline pre-training}, we can directly generalize PRoCD to large real graphs $\{ \mathcal{G}'\}$ (with frozen model parameters $\Theta^*$) and derive feasible CD results $\{ \mathcal{\bar C}' \}$ via only one FFP of the model and Algorithm~\ref{Alg:Post-Prep} (i.e., \textit{online generalization}).

Recent advances in graph pre-training techniques \cite{wu2021self} demonstrate that pre-training may provide good initialized model parameters for downstream tasks, which can be further improved via a task-specific fine-tuning procedure. To the best of our knowledge, most existing graph pre-training methods merely consider supervised tasks (e.g., node classification) but do not provide unsupervised tuning objectives for CD.
A straightforward fine-tuning strategy for CD is applying the modularity maximization objective (\ref{Eq:Mod-Obj}) to large real graphs $\{ \mathcal{G}' \}$ (e.g., gradient descent w.r.t. a relaxed version of (\ref{Eq:Mod-Max-Mat}) with large dense matrices $\{ {\bf{Q}}' \}$), which is usually intractable.
Analogous to fine-tuning, we introduce the \textit{online refinement} phase with a different design.
Instead of fine-tuning, we treat the CD result $\mathcal{\bar C}'$ derived in \textit{online generalization} as the initialization of a conventional efficient CD method (e.g., \textit{LPA} \cite{raghavan2007near}, \textit{InfoMap} \cite{rosvall2008maps}, and \textit{Locale} \cite{wang2020community} in our experiments) and run this method to derive a refined CD result $\mathcal{C}'$.

\begin{figure}[t]
  \centering
  \includegraphics[width=0.85\linewidth, trim=18 24 22 18,clip]{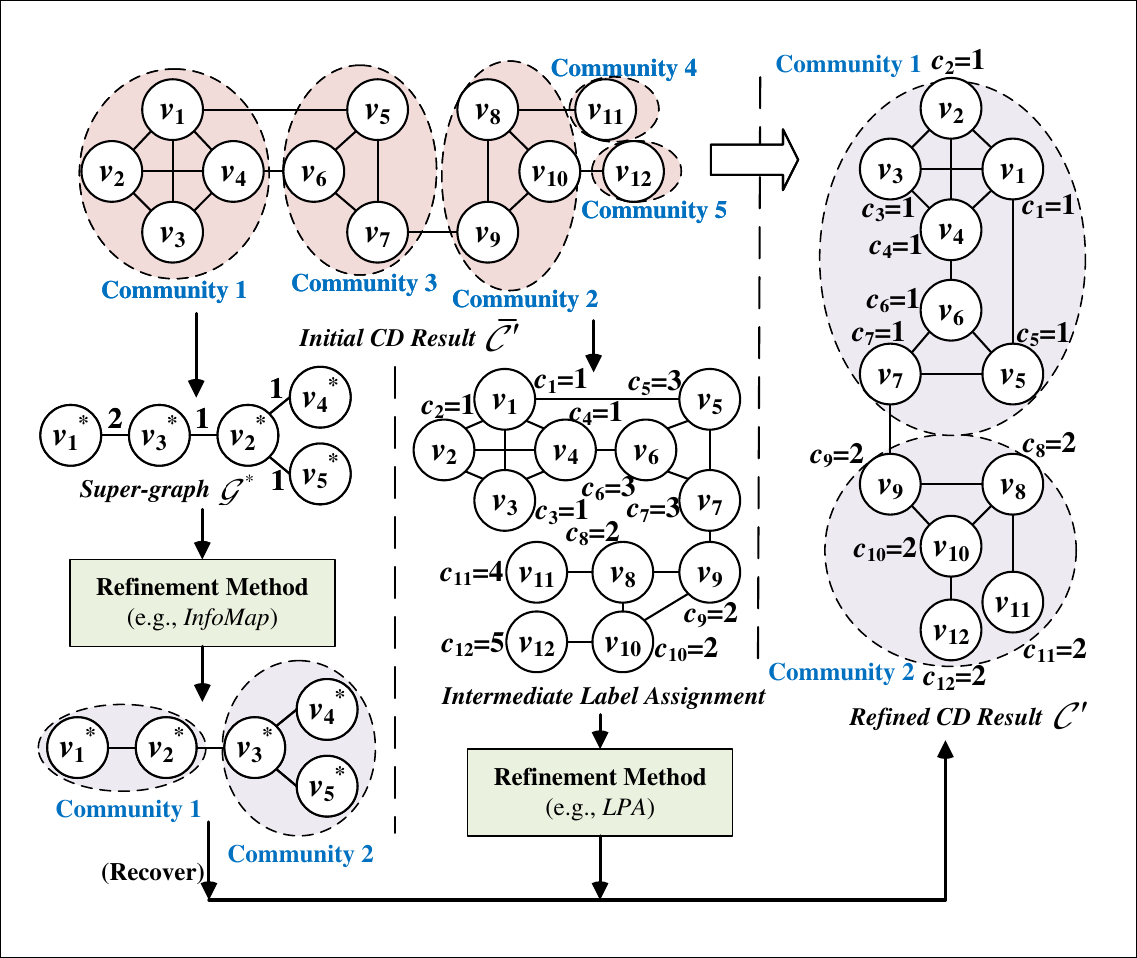}
  \caption{Illustration of the two strategies to construct the initialization of \textit{online refinement}.}\label{Fig:Rfn}
\end{figure}

As shown in Fig.~\ref{Fig:Rfn}, we adopt two strategies to construct the initialization. First, a super-graph $\mathcal{G}^*$ can be extracted based on $\mathcal{\bar C}'$, where we merge nodes in each community $\mathcal{\bar C}'_r \in \mathcal{\bar C}'$ as a super-node (e.g., $v^*_3$ w.r.t. $\mathcal{\bar C}'_3 = \{v_5, v_6, v_7\}$ in Fig.~\ref{Fig:Rfn}) and set the number of edges between communities as the weight of each super-edge (e.g., $2$ for $(v_1^*, v_3^*)$ in Fig.~\ref{Fig:Rfn}). We then use $\mathcal{G}^*$ as the input of an efficient method that can handle weighted graphs (e.g., \textit{InfoMap} and \textit{Locale}) to derive a CD result w.r.t. $\mathcal{G}^*$, which is further recovered to the result $\mathcal{C}'$ w.r.t. $\mathcal{G}'$.
Second, we also use $\mathcal{\bar C}'$ as the intermediate label assignment of a method that iteratively updates community labels (e.g., \textit{LPA}) and refines the label assignment until convergence (e.g., refining $c_{6}$ from $3$ to $1$ in Fig.~\ref{Fig:Rfn}).
Compared with running the refinement method on $\mathcal{G}'$ from scratch, \textit{online refinement} may be more efficient, because the constructed initialization reduces the number of nodes to be partitioned and iterations to update community labels.
It is also expected that PRoCD can transfer the capability of capturing community structures from the pre-training data to $\{ \mathcal{G}' \}$.

\begin{algorithm}[t]\footnotesize
\caption{\textit{Online Generalization \& Refinement}}
\label{Alg:Gen-Rfn}
\LinesNumbered
\KwIn{large real graph $\mathcal{G}'$ to be partitioned; saved model parameters $\Theta^*$}
\KwOut{CD results $\mathcal{C}'$ w.r.t. $\mathcal{G}'$}
get input feature ${\bf{X'}}$ w.r.t. $\mathcal{G}'$ via (\ref{Eq:Feat-Ext})\\
get a feasible CD result $\mathcal{\bar C}'$ w.r.t. $\mathcal{G}'$ via Algorithm~\ref{Alg:Post-Prep}\\
construct the initialization for refinement based on $\mathcal{\bar C}'$\\
get refined CD result $\mathcal{C}'$ via an efficient refinement method
\end{algorithm}

Algorithm~\ref{Alg:Gen-Rfn} concludes the aforementioned procedure.
The complexity of \textit{online generalization} (i.e., lines 1-2) is $\mathcal{O}((N + M)d^2)$, where $N$ and $M$ are the numbers of nodes and edges; $d$ is the embedding dimensionality. The complexity of \textit{online refinement} (i.e., lines 3-4) depends on the refinement method, which is usually efficient. We leave detailed complexity analysis in Appendix~\ref{App-Cmp}.

\begin{table}[]\footnotesize
\caption{Statistics of datasets.}\label{Tab:Data}
\begin{tabular}{l|l|l|lll|l}
\hline
\textbf{Datasets} & \textit{$N$} & \textit{$E$} & Min & Max & Avg Deg & Density \\ \hline
\textbf{Protein} \cite{stark2006biogrid} & 81,574 & 1,779,885 & 1 & 4,983 & 43.6 & 5e-4 \\
\textbf{ArXiv} \cite{wang2020microsoft} & 169,343 & 1,157,799 & 1 & 13,161 & 13.7 & 8e-5 \\
\textbf{DBLP} \cite{yang2012defining} & 317,080 & 1,049,866 & 1 & 343 & 6.6 & 2e-5 \\
\textbf{Amazon} \cite{yang2012defining} & 334,863 & 925,872 & 1 & 549 & 5.5 & 2e-5 \\
\textbf{Youtube} \cite{yang2012defining} & 1,134,890 & 2,987,624 & 1 & 28,754 & 5.3 & 5e-6 \\
\textbf{RoadCA} \cite{leskovec2009community} & 1,957,027 & 2,760,388 & 1 & 12 & 2.82 & 1e-6 \\ \hline
\end{tabular}
\end{table}

\begin{table*}[]\footnotesize
\caption{Evaluation results w.r.t. efficiency metric of inference time$\downarrow$ (sec) and quality metric of modularity$\uparrow$.}\label{Tab:Eva}
\begin{tabular}{p{1.45cm}|p{0.28cm}p{0.5cm}p{0.62cm}|p{0.38cm}p{0.5cm}p{0.62cm}|p{0.39cm}p{0.5cm}p{0.62cm}|p{0.39cm}p{0.52cm}p{0.67cm}|p{0.41cm}p{0.5cm}p{0.67cm}|p{0.5cm}p{0.5cm}l}
\hline
\multirow{2}{*}{} & \multicolumn{3}{c|}{\textbf{Protein}} & \multicolumn{3}{c|}{\textbf{ArXiv}} & \multicolumn{3}{c|}{\textbf{DBLP}} & \multicolumn{3}{c|}{\textbf{Amazon}} & \multicolumn{3}{c|}{\textbf{Youtube}} & \multicolumn{3}{c}{\textbf{RoadCA}} \\ \cline{2-19} 
 & \textit{K} & \textbf{Time}$\downarrow$ & \textbf{Mod}$\uparrow$ & \textit{K} & \textbf{Time}$\downarrow$ & \textbf{Mod}$\uparrow$ & \textit{K} & \textbf{Time}$\downarrow$ & \textbf{Mod}$\uparrow$ & \textit{K} & \textbf{Time}$\downarrow$ & \textbf{Mod}$\uparrow$ & \textit{K} & \textbf{Time}$\downarrow$ & \textbf{Mod}$\uparrow$ & \textit{K} & \textbf{Time}$\downarrow$ & \textbf{Mod}$\uparrow$ \\ \hline
MC-SBM & 170 & 1143.93 & 0.1463 & 303 & 1014.63 & 0.5205 & 430 & 646.91 & 0.7620 & 368 & 799.05 & 0.8860 & 186 & 6608.67 & 0.2027 & 324 & 8492.20 & 0.9491 \\
Par-SBM & 1494 & 39.18 & 0.4721 & 4936 & 40.26 & 0.0529 & 10854 & 28.04 & 0.2797 & 10813 & 39.40 & 0.3094 & 27948 & 236.61 & 0.0901 & 39987 & 149.10 & 0.5970 \\
GMod & 69 & 5840.49 & 0.6319 & 1317 & 6693.91 & 0.5797 & OOT & OOT & OOT & 1669 & 12729.36 & 0.8703 & OOT & OOT & OOT & OOT & OOT & OOT \\
Louvain & 60 & 46.23 & 0.6366 & 155 & 49.07 & 0.7059 & 209 & 83.50 & 0.8209 & 240 & 48.10 & 0.9262 & 6105 & 173.70 & 0.7218 & 376 & 408.25 & 0.9923 \\
RaftGP-C & 218 & 69.82 & 0.6124 & 213 & 79.11 & 0.6147 & 39 & 81.99 & 0.6594 & 53 & 87.66 & 0.7530 & 1 & 254.77 & 0.0000 & 256 & 470.05 & 0.8072 \\
RaftGP-M & 185 & 69.44 & 0.6199 & 194 & 80.10 & 0.6107 & 39 & 83.48 & 0.6644 & 54 & 88.24 & 0.7561 & 131 & 272.82 & 0.6659 & 256 & 469.14 & 0.8043 \\
\text{LouNE+{\tiny DBSCAN}} & 32 & 77.20 & 0.6362 & 170 & 214.55 & 0.7058 & 737 & 109.87 & 0.8065 & 252 & 64.45 & 0.9255 & 17908 & 671.52 & 0.6228 & 19081 & 236.46 & 0.9531 \\
\text{SktNE+{\tiny DBSCAN}} & 371 & 16.18 & 0.1808 & 388 & 542.92 & 0.3533 & 15341 & 45.60 & 0.5716 & 13283 & 49.85 & 0.7047 & 17799 & 485.72 & 0.1076 & 59726 & 1410.91 & 0.4073 \\
\text{ICD-C+{\tiny DBSCAN}} & 1277 & 89.30 & 0.4961 & 256 & 296.02 & 0.0011 & 3694 & 1795.04 & 0.1047 & 1865 & 2027.30 & 0.0278 & OOT & OOT & OOT & OOT & OOT & OOT \\
\text{ICD-M+{\tiny DBSCAN}} & 1091 & 88.47 & 0.5651 & 264 & 294.13 & 0.0019 & 3950 & 984.98 & 0.1156 & 2478 & 872.11 & 0.0430 & OOT & OOT & OOT & OOT & OOT & OOT \\ \hline
\textit{LPA} & 1919 & 132.32 & 0.5925 & 11467 & 113.30 & 0.3752 & 46630 & 146.22 & 0.6196 & 41693 & 76.49 & 0.7090 & 111887 & 370.15 & 0.5520 & 606249 & 245.54 & 0.5524 \\
\textbf{PRoCD}{\scriptsize w/~\textit{LPA}} & 1499 & \textbf{32.07} & \textbf{0.6046} & 10765 & \textbf{19.93} & \textbf{0.6188} & 38684 & \textbf{20.10} & \textbf{0.6336} & 34270 & \textbf{20.40} & \textbf{0.7240} & 93566 & \textbf{68.72} & \textbf{0.6150} & 391735 & \textbf{67.74} & \textbf{0.5765} \\
\textbf{~~Improve.} &  & \text{+75.8\%} & \text{+2.1\%} &  & \text{+82.4\%} & \text{+64.9\%} &  & \text{+86.3\%} & \text{+2.3\%} &  & \text{+73.3\%} & \text{+2.1\%} &  & \text{+81.4\%} & \text{+11.4\%} &  & \text{+72.4\%} & \text{+4.4\%} \\ \cline{2-19}
GC+\textit{LPA} & 1985 & 32.52 & 0.5790 & 12854 & 21.57 & 0.5290 & 44494 & 30.04 & 0.5906 & 41513 & 29.40 & 0.5723 & 124044 & 69.17 & 0.2710 & 442623 & 106.89 & 0.4838 \\ \hline
\textit{InfoMap} & 11 & 26.38 & 0.1972 & 69 & 19.69 & 0.6332 & 527 & 29.11 & 0.8195 & 13 & 28.72 & 0.8037 & 956 & 114.63 & 0.6959 & 3 & 259.44 & 0.6455 \\ 
\textbf{PRoCD}{\scriptsize w/~\textit{IM}} & 18 & \textbf{15.60} & \textbf{0.1996} & 42 & \textbf{17.08} & \textbf{0.6349} & 298 & \textbf{17.23} & \textbf{0.8222} & 417 & \textbf{16.66} & \textbf{0.9222} & 438 & \textbf{83.01} & {\underline{0.6811}} & 145 & \textbf{79.48} & \textbf{0.9889} \\
\textbf{~~Improve.} &  & \text{+40.9\%} & \text{+1.2\%} &  & \text{+13.3\%} & \text{+0.3\%} &  & \text{+40.8\%} & \text{+0.3\%} &  & \text{+42.0\%} & \text{+14.7\%} &  & \text{+27.6\%} & \text{-2.1\%} &  & \text{+69.4\%} & \text{+53.2\%} \\ \cline{2-19}
GC+\textit{IM} & 9 & 33.84 & 0.1318 & 48 & 24.28 & 0.4735 & 305 & 27.99 & 0.8086 & 4 & 28.01 & 0.4189 & 687 & 104.85 & 0.6919 & 3 & 129.74 & 0.6561 \\ \hline
\textit{Locale} & 51 & 43.04 & 0.6409 & 150 & 46.66 & 0.7159 & 298 & 40.08 & 0.8375 & 392 & 30.85 & 0.9343 & 3441 & 216.12 & 0.7353 & 410 & 172.48 & 0.9935 \\ 
\textbf{PRoCD}{\scriptsize w/~\textit{Lcl}} & 35 & \textbf{16.93} & {\underline{0.6362}} & 64 & \textbf{31.76} & {\underline{0.7133}} & 147 & \textbf{24.15} & \textbf{0.8376} & 173 & \textbf{18.94} & {\underline{0.9319}} & 1647 & \textbf{161.24} & {\underline{0.7315}} & 330 & \textbf{90.88} & {\underline{0.9930}} \\
\textbf{~~Improve.} & & \text{+60.7\%} & \text{-0.7\%} &  & \text{+31.9\%} & \text{-0.4\%} &  & \text{+39.8\%} & \text{+0.01\%} &  & \text{+38.6\%} & \text{-0.3\%} &  & \text{+25.4\%} & \text{-0.5\%} &  & \text{+47.3\%} & \text{-0.1\%} \\ \cline{2-19}
GC+\textit{Lcl} & 33 & 44.35 & 0.6343 & 150 & 45.17 & 0.6988 & 196 & 36.87 & 0.8254 & 344 & 30.09 & 0.9220 & 3228 & 194.49 & 0.7315 & 327 & 130.57 & 0.9928 \\ \hline
\end{tabular}
\end{table*}

\begin{table}[]\footnotesize
\caption{Detailed inference time (sec) of PRoCD.}\label{Tab:Time}
\begin{tabular}{c|l|cccl}
\hline
\multicolumn{1}{l|}{\textbf{Datasets}} & \textbf{Methods} & \multicolumn{1}{l}{\textbf{Feat}} & \multicolumn{1}{l}{\textbf{FFP}} & \multicolumn{1}{l}{\textbf{Init}} & \textbf{Rfn} \\ \hline
\multirow{3}{*}{\textbf{Protein}} & \textbf{PRoCD} w/ \textit{LPA} & \multirow{3}{*}{2.84} & \multirow{3}{*}{4.42} & \multirow{3}{*}{6.50} & 18.31 \\
 & \textbf{PRoCD} w/ \textit{InfoMap} &  &  &  & 1.84 \\
 & \textbf{PRoCD} w/ \textit{Locale} &  &  &  & 3.17 \\ \hline
\multirow{3}{*}{\textbf{ArXiv}} & \textbf{PRoCD} w/ \textit{LPA} & \multirow{3}{*}{1.49} & \multirow{3}{*}{0.99} & \multirow{3}{*}{4.50} & 12.96 \\
 & \textbf{PRoCD} w/ \textit{InfoMap} &  &  &  & 10.11 \\
 & \textbf{PRoCD} w/ \textit{Locale} &  &  &  & 24.79 \\ \hline
\multirow{3}{*}{\textbf{DBLP}} & \textbf{PRoCD} w/ \textit{LPA} & \multirow{3}{*}{1.97} & \multirow{3}{*}{1.28} & \multirow{3}{*}{4.19} & 12.66 \\
 & \textbf{PRoCD} w/ \textit{InfoMap} &  &  &  & 9.79 \\
 & \textbf{PRoCD} w/ \textit{Locale} &  &  &  & 16.71 \\ \hline
\multirow{3}{*}{\textbf{Amazon}} & \textbf{PRoCD} w/ \textit{LPA} & \multirow{3}{*}{2.08} & \multirow{3}{*}{1.99} & \multirow{3}{*}{3.81} & 12.51 \\
 & \textbf{PRoCD} w/ \textit{InfoMap} &  &  &  & 8.78 \\
 & \textbf{PRoCD} w/ \textit{Locale} &  &  &  & 11.06 \\ \hline
\multirow{3}{*}{\textbf{Youtube}} & \textbf{PRoCD} w/ \textit{LPA} & \multirow{3}{*}{5.81} & \multirow{3}{*}{2.75} & \multirow{3}{*}{15.13} & 45.02 \\
 & \textbf{PRoCD} w/ \textit{InfoMap} &  &  &  & 59.32 \\
 & \textbf{PRoCD} w/ \textit{Locale} &  &  &  & 137.54 \\ \hline
\multirow{3}{*}{\textbf{RoadCA}} & \textbf{PRoCD} w/ \textit{LPA} & \multirow{3}{*}{8.10} & \multirow{3}{*}{2.34} & \multirow{3}{*}{13.14} & 44.16 \\
 & \textbf{PRoCD} w/ \textit{InfoMap} &  &  &  & 55.90 \\
 & \textbf{PRoCD} w/ \textit{Locale} &  &  &  & 67.31 \\ \hline
\end{tabular}
\end{table}

\begin{table}[]\footnotesize
\caption{Detailed inference time (sec) of graph GC baselines.}\label{Tab:Time-GC}
\begin{tabular}{c|l|cl}
\hline
\multicolumn{1}{l|}{\textbf{Datasets}} & \textbf{Methods} & \multicolumn{1}{l}{\textbf{GC}} & \textbf{Rfn} \\ \hline
\multirow{3}{*}{\textbf{Protein}} & GC+\textit{LPA} & \multirow{3}{*}{15.20} & 17.32 \\
 & GC+\textit{InfoMap} &  & 18.64 \\
 & GC+\textit{Locale} &  & 29.15 \\ \hline
\multirow{3}{*}{\textbf{ArXiv}} & GC+\textit{LPA} & \multirow{3}{*}{10.20} & 11.37 \\
 & GC+\textit{InfoMap} &  & 14.08 \\
 & GC+\textit{Locale} &  & 34.97 \\ \hline
\multirow{3}{*}{\textbf{DBLP}} & GC+\textit{LPA} & \multirow{3}{*}{17.84} & 12.20 \\
 & GC+\textit{InfoMap} &  & 10.15 \\
 & GC+\textit{Locale} &  & 19.03 \\ \hline
\multirow{3}{*}{\textbf{Amazon}} & GC+\textit{LPA} & \multirow{3}{*}{17.19} & 12.21 \\
 & GC+\textit{InfoMap} &  & 10.82 \\
 & GC+\textit{Locale} &  & 12.90 \\ \hline
\multirow{3}{*}{\textbf{Youtube}} & GC+\textit{LPA} & \multirow{3}{*}{31.84} & 37.32 \\
 & GC+\textit{InfoMap} &  & 73.01 \\
 & GC+\textit{Locale} &  & 162.65 \\ \hline
\multirow{3}{*}{\textbf{RoadCA}} & GC+\textit{LPA} & \multirow{3}{*}{71.33} & 35.56 \\
 & GC+\textit{InfoMap} &  & 58.41 \\
 & GC+\textit{Locale} &  & 59.24 \\ \hline
\end{tabular}
\end{table}

\begin{table}[]\footnotesize
\caption{Detailed inference time (s) of embedding baselines.}\label{Tab:Time-Emb}
\begin{tabular}{c|l|ll}
\hline
\multicolumn{1}{l|}{\textbf{Datasets}} & \textbf{Methods} & \textbf{Emb} & \textbf{Clus} \\ \hline
\multirow{4}{*}{\textbf{Protein}} & LouvainNE+DBSCAN & 4.13 & 73.07 \\
 & SketchNE+DBSCAN & 4.30 & 11.88 \\
 & ICD-C+DBSCAN & 6.99 & 82.31 \\
 & ICD-M+DBSCAN & 12.49 & 75.97 \\ \hline
\multirow{4}{*}{\textbf{ArXiv}} & LouvainNE+DBSCAN & 6.86 & 207.41 \\
 & SketchNE+DBSCAN & 5.79 & 537.12 \\
 & ICD-C+DBSCAN & 21.59 & 274.43 \\
 & ICD-M+DBSCAN & 19.57 & 274.55 \\ \hline
\multirow{4}{*}{\textbf{DBLP}} & LouvainNE+DBSCAN & 10.52 & 99.35 \\
 & SketchNE+DBSCAN & 7.71 & 37.90 \\
 & ICD-C+DBSCAN & 5.66 & 1789.38 \\
 & ICD-M+DBSCAN & 11.53 & 973.45 \\ \hline
\multirow{4}{*}{\textbf{Amazon}} & LouvainNE+DBSCAN & 10.79 & 53.67 \\
 & SketchNE+DBSCAN & 8.05 & 41.79 \\
 & ICD-C+DBSCAN & 16.26 & 2011.04 \\
 & ICD-M+DBSCAN & 23.22 & 848.89 \\ \hline
\multirow{4}{*}{\textbf{Youtube}} & LouvainNE+DBSCAN & 34.98 & 635.55 \\
 & SketchNE+DBSCAN & 19.40 & 466.32 \\
 & ICD-C+DBSCAN & 10.43 & OOT \\
 & ICD-M+DBSCAN & 7.66 & OOT \\ \hline
\multirow{4}{*}{\textbf{RoadCA}} & LouvainNE+DBSCAN & 56.30 & 180.15 \\
 & SketchNE+DBSCAN & 25.11 & 1385.80 \\
 & ICD-C+DBSCAN & 14.49 & OOT \\
 & ICD-M+DBSCAN & 12.90 & OOT \\ \hline
\end{tabular}
\end{table}

\section{Experiments}\label{Sec:Exp}

In this section, we elaborate on our experiments, including the experiment setup, evaluation results, ablation study, and parameter analysis.
We will make the datasets and demo code public at https://github.com/KuroginQin/PRoCD.

\subsection{Experiment Setup}

\textbf{Datasets}.
We evaluated the inference efficiency and quality of our \textbf{PRoCD} method on $6$ public datasets with statistics depicted in Table~\ref{Tab:Data}, where $N$ and $M$ are the numbers of nodes and edges.
Note that these datasets do not provide ground-truth about the community assignment and the number of communities $K$.
Some more details regarding the datasets can be found in Appendix~\ref{App-Data}.

\textbf{Baselines}.
We compared \textbf{PRoCD} over $11$ baselines.

(\romannumeral1) \textit{MC-SBM} \cite{peixoto2014efficient}, (\romannumeral2) \textit{Par-SBM} \cite{peng2015scalable}, (\romannumeral3) \textit{GMod} \cite{clauset2004finding}, (\romannumeral4) \textit{LPA} \cite{raghavan2007near}, (\romannumeral5) \textit{InfoMap} \cite{rosvall2008maps}, (\romannumeral6) \textit{Louvain} \cite{blondel2008fast}, and (\romannumeral7) \textit{Locale} \cite{wang2020community}, and (\romannumeral8) \textit{RaftGP} \cite{gao2023raftgp} are efficient end-to-end methods that can tackle $K$-agnostic CD.
Note that some other approaches whose designs rely on a pre-set $K$ (e.g., \textit{FastCom} \cite{clauset2004finding}, \textit{GraClus} \cite{dhillon2007weighted}, \textit{ClusterNet} \cite{wilder2019end}, \textit{LGNN} \cite{chen2019supervised}, and \textit{DMoN} \cite{tsitsulin2023graph}) could not be included in our experiments.

In addition, we extended (\romannumeral9) \textit{LouvainNE} \cite{bhowmick2020louvainne}, (\romannumeral10) \textit{SketchNE} \cite{xie2023sketchne}, and (\romannumeral11) \textit{ICD} \cite{qin2023towards}, which are state-of-the-art efficient graph embedding methods, to $K$-agnostic CD by combining the derived embeddings with DBSCAN \cite{schubert2017dbscan}, an efficient clustering algorithm that does not require $K$.
In particular, \textit{LouvainNE} and \textit{ICD} are claimed to be community-preserving methods.
\textit{RaftGP-C}, \textit{RaftGP-M}, \textit{ICD-C}, and \textit{ICD-M} denote different variants of \textit{RaftGP} and \textit{ICD}.

As a demonstration, we adopted \textit{LPA}, \textit{InfoMap}, and \textit{Locale} (an advanced extension of \textit{Louvain}) for the \textit{online refinement} of \textbf{PRoCD}, forming three variants of our method.
In Fig.~\ref{Fig:Rfn}, the first strategy of constructing the refinement initialization has a motivation similar to graph coarsening (GC) techniques \cite{chen2022graph}, which merge a large graph $\mathcal{G}'$ into a super-graph $\mathcal{G}^*$ (i.e., reducing the number of nodes to be partitioned) via a heuristic strategy. Whereas, \textbf{PRoCD} constructs $\mathcal{G}^*$ based on the output of a pre-trained model.
To highlight the superiority of \textit{offline pre-training} and inductive inference beyond GC, we introduced another baseline that provides the initialization (with the same number of super-nodes as \textbf{PRoCD}) for \textit{online refinement} using the efficient GC strategy proposed in \cite{liang2021mile}.

\textit{ICD} is an inductive baseline with a paradigm similar to \textbf{PRoCD}, including \textit{offline (pre-)training} on historical graphs and \textit{online generalization} to new graphs. However, there is no \textit{online refinement} in \textit{ICD}.
We conducted the \textit{offline pre-training} of \textit{ICD} and \textbf{PRoCD} on the generated synthetic graphs as described in Section~\ref{Sec:PTN-Data-Gen} and generalized them to each dataset for online inference.
For the rest baselines, we had to run them from scratch on each dataset.

\textbf{Evaluation Criteria}. Following the evaluation protocol described in Section~\ref{Sec:Prob}, we adopted modularity and inference time (sec) as the quality and efficiency metrics. Moreover, we define that a method encounters the out-of-time (OOT) exception if it fails to derive a feasible CD result within $2 \times 10^4$ seconds.

Due to space limits, we elaborate on other details of experiment setup (e.g., parameter settings) in Appendix~\ref{App-Exp-Set}.

\subsection{Quantitative Evaluation \& Discussions}

The average evaluation results are reported in Table~\ref{Tab:Eva}.
Besides the total inference time, we also report the time w.r.t. each inference phase of PRoCD in Table~\ref{Tab:Time}, where `Feat', `FFP', `Init', and `Rfn' denote the time of (\romannumeral1) feature extraction, (\romannumeral2) one FFP, (\romannumeral3) initial result derivation, and (\romannumeral4) online refinement. Analysis about the inference time of GC and embedding baselines is given in Tables~\ref{Tab:Time-GC} and \ref{Tab:Time-Emb}, where `GC', `Rfn', `Emb', and `Clus' denote the time of (\romannumeral1) graph coarsening, (\romannumeral2) online refinement, (\romannumeral3) embedding derivation, and (\romannumeral4) downstream clustering (i.e., via DBSCAN).

In Table~\ref{Tab:Eva}, three variants of PRoCD achieve significant improvement of efficiency w.r.t. the refinement methods while the quality degradation is less than $2\%$. In some cases, PRoCD can even obtain improvement for both aspects.
Compared with all the baselines, PRoCD can ensure the best efficiency on all the datasets and is in groups with the best quality.
In summary, we believe that the proposed method provides a possible way to obtain a better trade-off between the quality and efficiency in CD.

In most cases, PRoCD outperforms GC baselines in both aspects, which validates the superiority of offline pre-training and inductive inference beyond heuristic GC.
The extensions of some embedding baselines (e.g., \textit{SketchNE} and \textit{ICD} combined with DBSCAN) suffer from low quality and efficiency compared with other end-to-end approaches (e.g., \textit{Louvain} and \textit{Locale}). It implies that the task-specific module (e.g., binary node pair classifier in PRoCD) is essential for the embedding framework to ensure the quality and efficiency in $K$-agnostic CD, which is seldom considered in related studies.

According to Table~\ref{Tab:Time}, online refinement is the major bottleneck in the inference of PRoCD, with the runtime depending on concrete refinement methods.
Compared with running a refinement method from scratch, online refinement has much lower runtime. It validates our motivation that constructing the initialization (as described in Fig.~\ref{Fig:Rfn}) can reduce the number of nodes to be partitioned and iterations to update label assignments for refinement methods.
To construct the initialization of online refinement, GC is not as efficient as the online generalization of PRoCD (i.e., feature extraction, one FFP, and initial result derivation) according to Table~\ref{Tab:Time-GC}.
In Table~\ref{Tab:Time-Emb}, the downstream clustering (i.e., DBSCAN to derive a feasible result for $K$-agnostic CD) is time-consuming, which results in low efficiency of the extended embedding baselines, although their embedding derivation phases are efficient.

\begin{table}[]\footnotesize
\caption{Ablation studies on DBLP and Amazon in terms of modularity$\uparrow$ and inference time$\downarrow$ (sec).}\label{Tab:Abl}
\begin{tabular}{l|lll|lll}
\hline
\multirow{2}{*}{} & \multicolumn{3}{c|}{\textbf{DBLP}} & \multicolumn{3}{c}{\textbf{Amazon}} \\ \cline{2-7} 
 & \textit{K} & \textbf{Mod}$\uparrow$ & \textbf{Time}$\downarrow$ & \textit{K} & \textbf{Mod}$\uparrow$ & \textbf{Time}$\downarrow$ \\ \hline
(\romannumeral1)w/o Rfn & 106553 & 0.4394 & 7.44 & 112408 & 0.5620 & 7.88 \\ \hline
\textbf{PRoCD}~w/~\textit{LPA} & 38684 & 0.6336 & 20.10 & 34270 & 0.7240 & 20.40 \\ \cline{2-7} 
(\romannumeral2)w/o Ptn & 1 & 0.0000 & 9.75 & 1 & 0.0000 & 8.81 \\
(\romannumeral3)w/o Feat (\ref{Eq:Feat-Ext}) & 537 & 0.0625 & 18.01 & 1 & 0.0000 & 6.29 \\
(\romannumeral4)w/o EmbEnc (\ref{Eq:Emb-Enc}) & 58401 & 0.5406 & 18.26 & 53127 & 0.5090 & 17.08 \\
(\romannumeral5)w/o BinClf (\ref{Eq:Bin-Clf}) & 48 & 0.0009 & 6.43 & 23 & 0.0008 & 6.31 \\
(\romannumeral6)w/o $\mathcal{L}_{\rm{BCE}}$ (\ref{Eq:BCE-Obj}) & 22496 & 0.5657 & 17.84 & 16378 & 0.6580 & 20.19 \\
(\romannumeral7) w/o $\mathcal{L}_{\rm{MOD}}$ (\ref{Eq:Mod-Obj}) & 39792 & 0.6261 & 20.12 & 34853 & 0.7188 & 19.31 \\ \hline
\textbf{PRoCD}~w/~\textit{IM} & 298 & 0.8222 & 17.23 & 417 & 0.9222 & 16.66 \\ \cline{2-7} 
(\romannumeral2)w/o Ptn & 1 & 0.0000 & 9.75 & 1 & 0.0000 & 8.81 \\
(\romannumeral3)w/o Feat (\ref{Eq:Feat-Ext}) & 2 & 0.0001 & 5.67 & 1 & 0.0000 & 5.13 \\
(\romannumeral4)w/o EmbEnc (\ref{Eq:Emb-Enc}) & 516 & 0.8182 & 26.55 & 6 & 0.3869 & 39.27 \\
(\romannumeral5)w/o BinClf (\ref{Eq:Bin-Clf}) & 48 & 0.0009 & 5.88 & 23 & 0.0008 & 5.68 \\
(\romannumeral6)w/o $\mathcal{L}_{\rm{BCE}}$ (\ref{Eq:BCE-Obj}) & 156 & 0.2895 & 12.93 & 356 & 0.4872 & 9.91 \\
(\romannumeral7)w/o $\mathcal{L}_{\rm{MOD}}$ (\ref{Eq:Mod-Obj}) & 306 & 0.8200 & 19.22 & 2 & 0.2638 & 22.06 \\ \hline
\textbf{PRoCD}~w/~\textit{Lcl} & 147 & 0.8376 & 24.15 & 173 & 0.9319 & 18.94 \\ \cline{2-7} 
(\romannumeral2)w/o Ptn & 1 & 0.0000 & 9.75 & 1 & 0.0000 & 8.81 \\
(\romannumeral3)w/o Feat (\ref{Eq:Feat-Ext}) & 551 & 0.0647 & 5.66 & 1 & 0.0000 & 5.15 \\
(\romannumeral4)w/o EmbEnc (\ref{Eq:Emb-Enc}) & 216 & 0.8368 & 42.20 & 307 & 0.9228 & 35.97 \\
(\romannumeral5)w/o BinClf (\ref{Eq:Bin-Clf}) & 48 & 0.0009 & 6.66 & 23 & 0.0008 & 6.43 \\
(\romannumeral6)w/o $\mathcal{L}_{\rm{BCE}}$ (\ref{Eq:BCE-Obj}) & 305 & 0.7460 & 18.04 & 776 & 0.7448 & 19.23 \\
(\romannumeral7)w/o $\mathcal{L}_{\rm{MOD}}$ (\ref{Eq:Mod-Obj}) & 143 & 0.8365 & 25.97 & 210 & 0.9228 & 17.76 \\ \hline
\end{tabular}
\end{table}

\subsection{Ablation Study}
For our PRoCD method, we verified the effectiveness of (\romannumeral1) online refinement, (\romannumeral2) offline pre-training, (\romannumeral3) feature extraction module in (\ref{Eq:Feat-Ext}), (\romannumeral4) embedding encoder in (\ref{Eq:Emb-Enc}), (\romannumeral5) binary node pair classifier in (\ref{Eq:Bin-Clf}), (\romannumeral6) cross-entropy objective $\mathcal{L}_{\rm{BCE}}$, and (\romannumeral7) modularity maximization objective $\mathcal{L}_{\rm{MOD}}$ by removing corresponding components from the original model. In case (\romannumeral3), we let ${\bf{X}}$ be the one-hot encodings of node degree, which is a standard feature extraction strategy for GNNs when attributes are unavailable. We directly used ${\bf{X}}$ as the derived embeddings in case (\romannumeral4), while we adopted the naive binary classifier (\ref{Eq:Naive-Bin-Clf}) to replace our design (\ref{Eq:Bin-Clf}) in case (\romannumeral5). The average ablation study results on DBLP and Amazon are reported in Table~\ref{Tab:Abl}, where there are quality declines in all the cases. In some extreme cases (e.g., the one without pre-training), the model mistakenly partitions all the nodes into one community, resulting in a modularity value of $0.0000$.
In summary, all the considered procedures and components are essential to ensure the quality of PRoCD.

\begin{figure}[]
 \begin{minipage}{0.49\linewidth}
 \subfigure[DBLP, w/ \textit{LPA}]{
  \includegraphics[width=\textwidth,trim=42 0 40 0,clip]{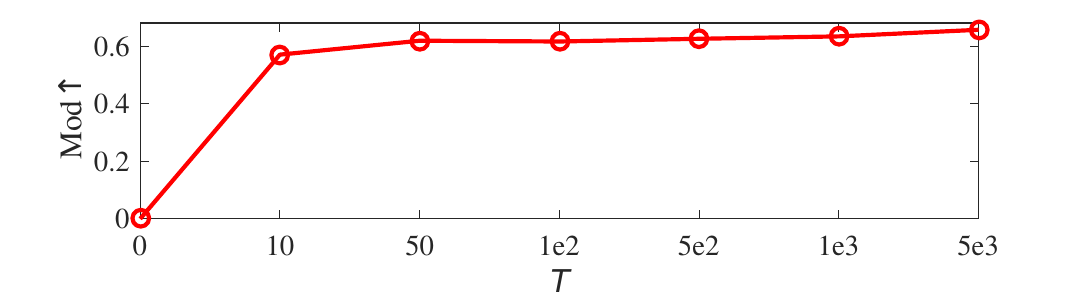}
  }
 \end{minipage}
 \begin{minipage}{0.49\linewidth}
 \subfigure[Amazon, w/ \textit{LPA}]{
  \includegraphics[width=\textwidth,trim=42 0 40 0,clip]{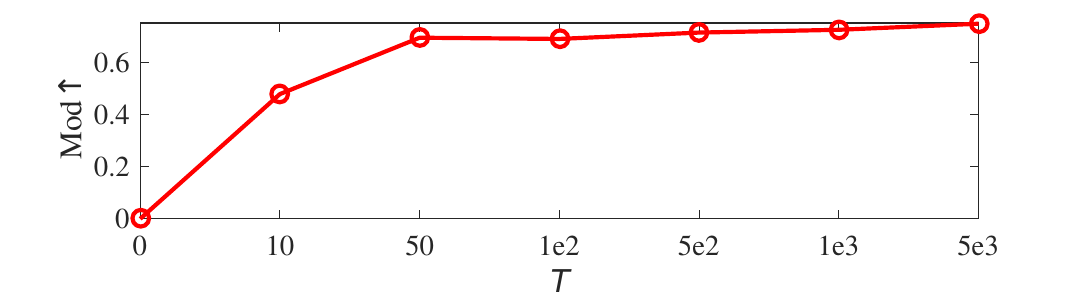}
  }
 \end{minipage}
 \begin{minipage}{0.49\linewidth}
 \subfigure[DBLP, w/ \textit{InfoMap}]{
  \includegraphics[width=\textwidth,trim=42 0 40 0,clip]{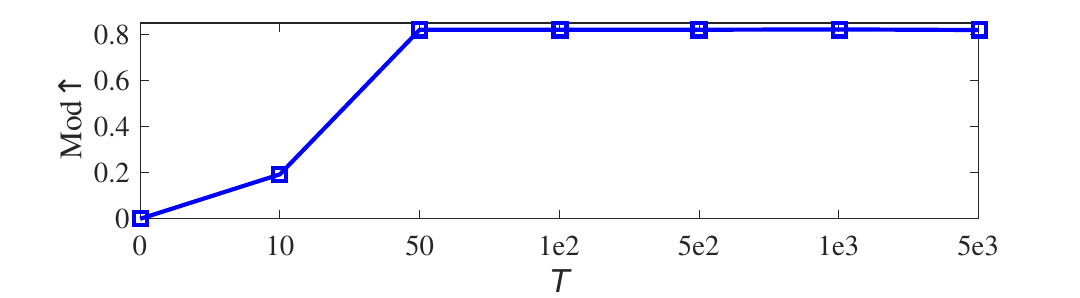}
  }
 \end{minipage}
 \begin{minipage}{0.49\linewidth}
 \subfigure[Amazon, w/ \textit{InfoMap}]{
  \includegraphics[width=\textwidth,trim=42 0 40 0,clip]{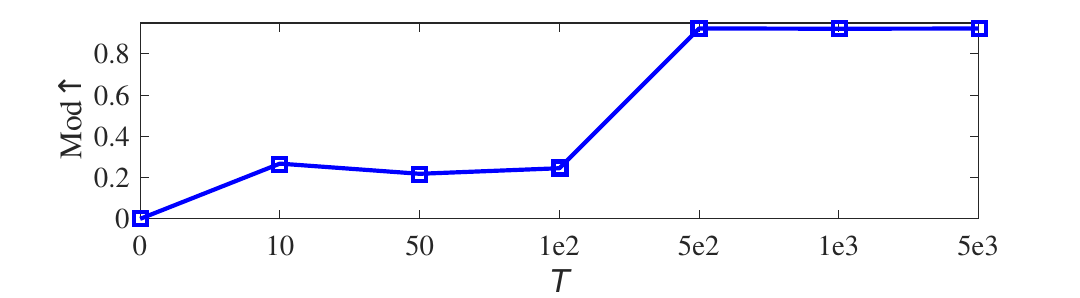}
  }
 \end{minipage}
 \begin{minipage}{0.49\linewidth}
 \subfigure[DBLP, w/ \textit{Locale}]{
  \includegraphics[width=\textwidth,trim=42 0 40 0,clip]{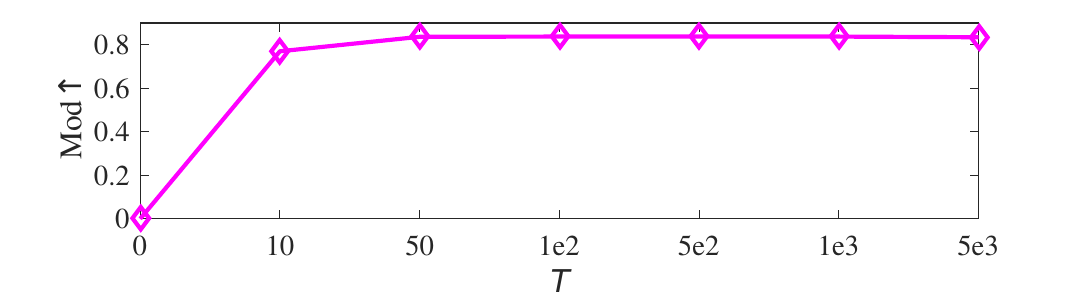}
  }
 \end{minipage}
 \begin{minipage}{0.49\linewidth}
 \subfigure[Amazon, w/ \textit{Locale}]{
  \includegraphics[width=\textwidth,trim=42 0 40 0,clip]{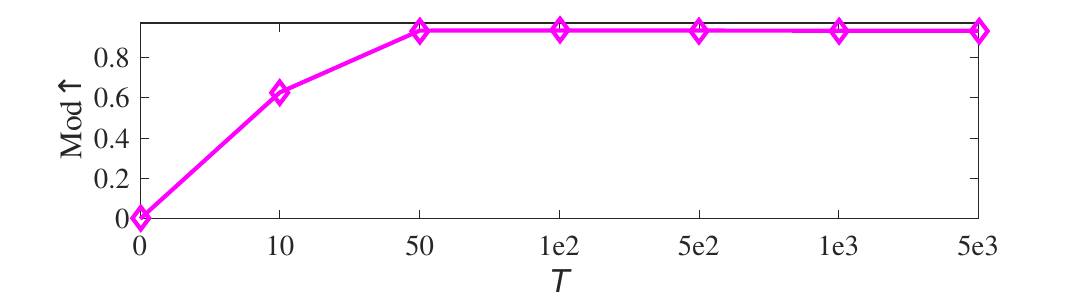}
  }
 \end{minipage}
 \caption{Ablation study w.r.t. the number of pre-training graphs $T$ on DBLP and Amazon in terms of modularity$\uparrow$.
 }\label{Fig:Abl-T}
\end{figure}

To further verify the significance of offline pre-training in our PRoCD method, we conducted additional ablation study by setting the number of pre-training graphs $T \in \{ 0, 10, 50, 1e2, 5e2, 1e3, 5e3 \}$. The corresponding results on DBLP and Amazon in terms of modularity are reported in Fig.~\ref{Fig:Abl-T}. Compared with our standard setting (i.e., $T = 1e3$), there are significant quality declines for all the variants of PRoCD when there are too few pre-training graphs (e.g., $T < 50$). It implies that \textit{the offline pre-training (with enough synthetic graphs) is essential to ensure the high inference quality of PRoCD}.

\begin{figure}[]
\centering
 \begin{minipage}{0.30\linewidth}
 \subfigure[DBLP, w/ \textit{LPA}]{
  \includegraphics[width=\textwidth,trim=10 5 0 15,clip]{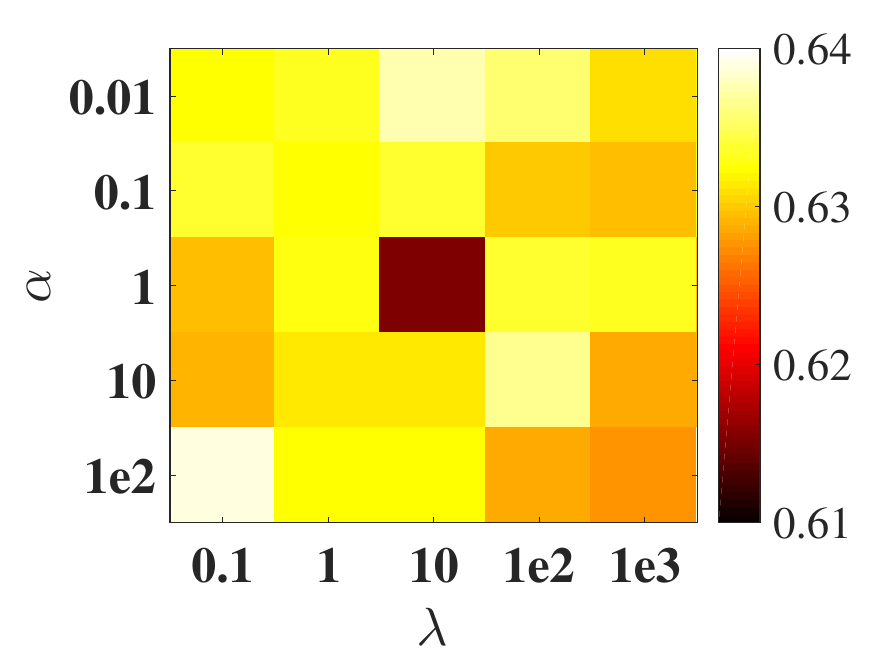}
  }
 \end{minipage}
 \begin{minipage}{0.30\linewidth}
 \subfigure[DBLP, w/ \textit{InfoMap}]{
  \includegraphics[width=\textwidth,trim=10 5 0 15,clip]{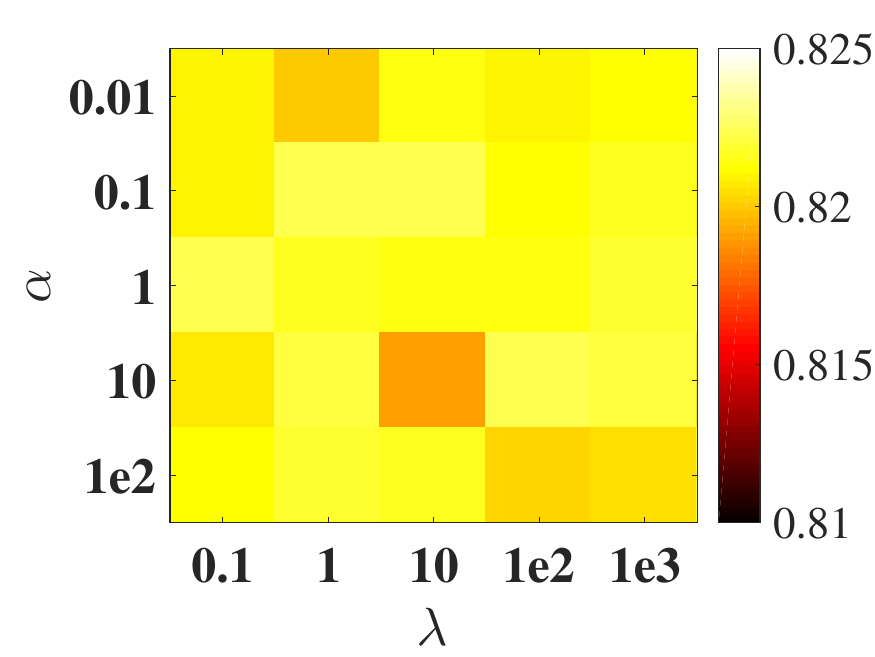}
  }
 \end{minipage}
 \begin{minipage}{0.30\linewidth}
 \subfigure[DBLP, w/ \textit{Locale}]{
  \includegraphics[width=\textwidth,trim=10 5 0 15,clip]{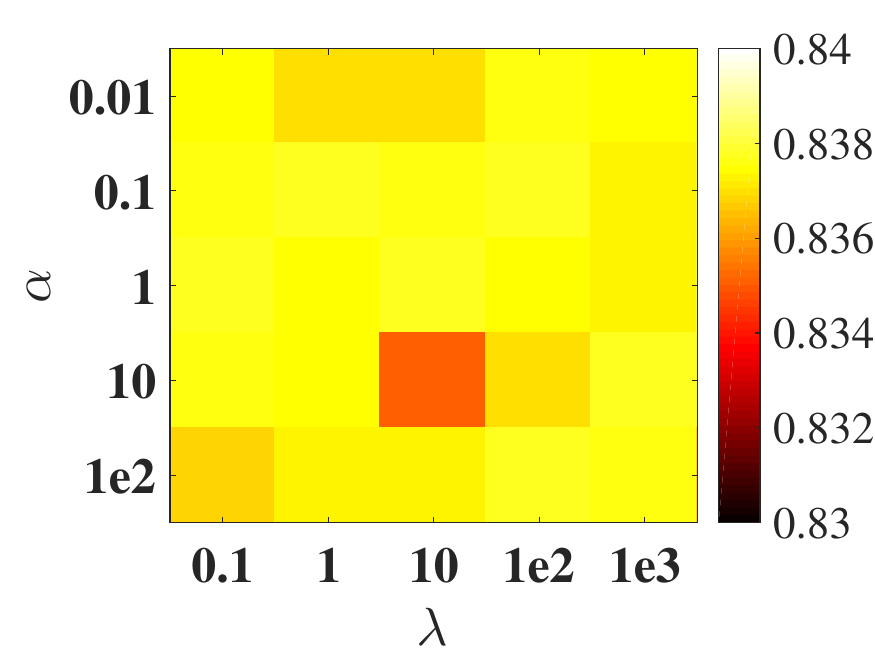}
  }
 \end{minipage}
 \begin{minipage}{0.30\linewidth}
 \subfigure[Amazon, w/ \textit{LPA}]{
  \includegraphics[width=\textwidth,trim=10 5 0 15,clip]{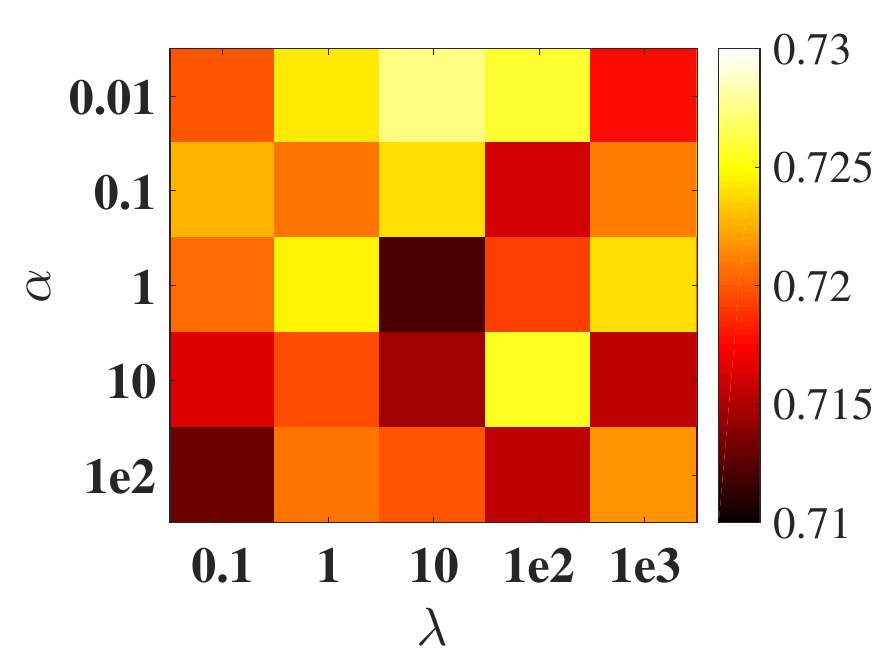}
  }
 \end{minipage}
 \begin{minipage}{0.30\linewidth}
 \subfigure[Amazon, w/ \textit{InfoMap}]{
  \includegraphics[width=\textwidth,trim=10 5 0 15,clip]{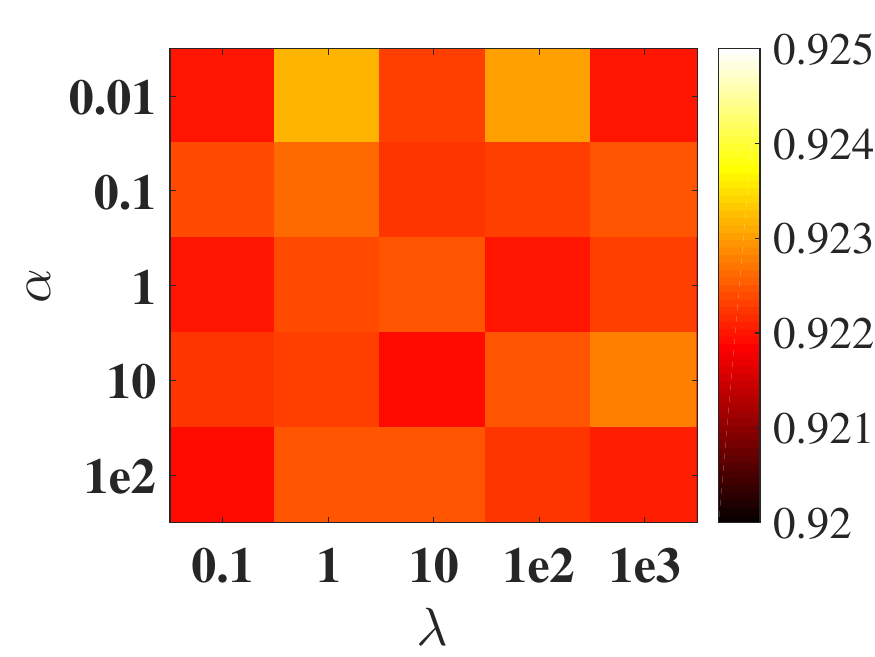}
  }
 \end{minipage}
 \begin{minipage}{0.30\linewidth}
 \subfigure[Amazon, w/ \textit{Locale}]{
  \includegraphics[width=\textwidth,trim=10 5 0 15,clip]{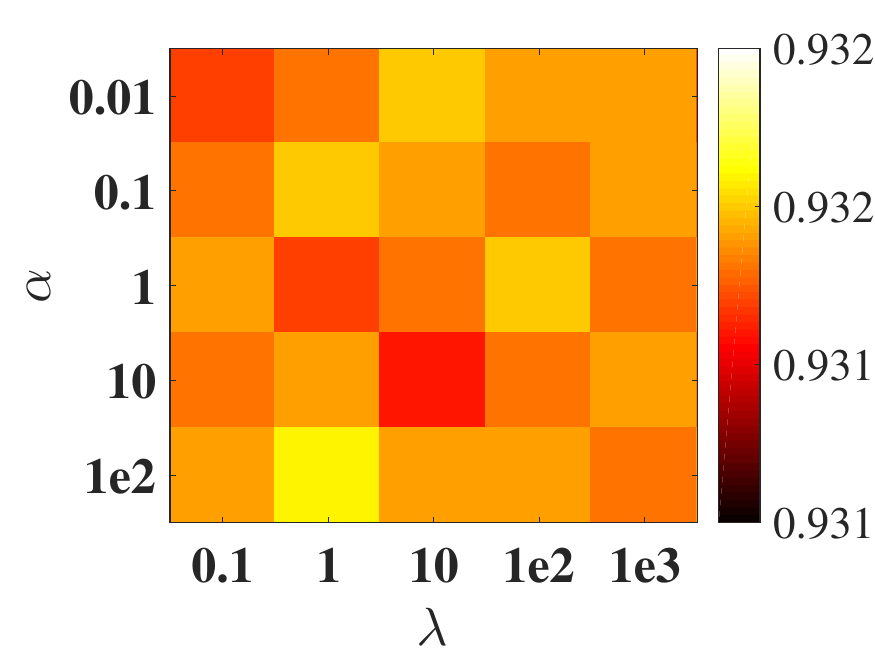}
  }
 \end{minipage}
 \caption{Parameter analysis w.r.t. $\alpha$ and $\lambda$ on DBLP and Amazon in terms of modularity$\uparrow$.
 }\label{Fig:Param}
\end{figure}

\subsection{Parameter Analysis}

We also tested the effects of $\{ \alpha, \lambda \}$ in the pre-training objective (\ref{Eq:PTN-Obj}) by adjusting $\alpha \in \{ 0.01, 0.1, 1, 10, 100\}$ and $\lambda \in \{ 0.1, 1, 10, 100, 1000\}$. The example parameter analysis results on DBLP and Amazon in terms of modularity are shown in Fig.~\ref{Fig:Param}, where different settings of $\{ \alpha, \lambda \}$ would not significantly affect the quality of PRoCD. The influence of $\{ \alpha, \lambda \}$ may also differ from refinement methods. For instance, \textit{LPA} is more sensitive to the parameter setting, compared with \textit{InfoMap} and \textit{Locale}.

Due to space limit, we leave further parameter analysis about the number of sampled node pair $n_S$ (see Algorithm~\ref{Alg:Post-Prep}) in Appendix~\ref{App-Exp-Res}.

\section{Conclusion}\label{Sec:Con}
In this paper, we explored the potential of DGL to achieve a better trade-off between the quality and efficiency of $K$-agnostic CD.
By reformulating this task as the binary node pair classification, a simple yet effective PRoCD method was proposed.
It follows a novel \textit{pre-training \& refinement} paradigm, including the (\romannumeral1) \textit{offline pre-training} on small synthetic graphs with various topology properties and high-quality ground-truth as well as the (\romannumeral2) \textit{online generalization} to and (\romannumeral3) \textit{refinement} on large real graphs without additional model optimization.
Extensive experiments demonstrate that PRoCD, combined with different refinement methods, can achieve higher inference efficiency without significant quality degradation on public real graphs with various scales.
Some possible future research directions are summarized in Appendix~\ref{App:Future}.

\begin{acks}
This research has been made possible by funding support provided to Dit-Yan Yeung by the Research Grants Council of Hong Kong under the Research Impact Fund project R6003-21.
\end{acks}


\bibliographystyle{ACM-Reference-Format}

\appendix

\section{Generation of Pre-training Data}\label{App-Syn-Alg}

\begin{table*}[]\footnotesize
\caption{Parameter settings of the synthetic graph generator.}\label{Tab:Syn-Gen}
\vspace{-0.3cm}
\begin{tabular}{l|ll|ll|l|l|l}
\hline
$T$ & $N$ & $K$ & ${\deg}_{\min}$ & ${\deg}_{\max}$ & $\gamma$ & $\mu$ & $\rho$ \\ \hline
1e3 & \textit{I}(2e3,~5e3) & \textit{I}(2,~1e3) & min\{5,~$\left\lceil {N/(4K)} \right\rceil$ \} & min\{5e2,~$\left\lceil {N/K} \right\rceil $\} & \textit{F}(2,~3.5) & \textit{F}(2.5,~5) & \textit{F}(1,~3) \\ \hline
\end{tabular}
\vspace{-0.3cm}
\end{table*}

\begin{table*}[]\footnotesize
\caption{Statistics of the generated synthetic pre-training graphs.}\label{Tab:Stat-Syn-Data}
\vspace{-0.3cm}
\begin{tabular}{lll|lll|lll|l|l|lll}
\hline
Min & Max & Avg $N$ & Min & Max & Avg $E$ & Min & Max & Avg $K$ & Avg ${\deg}_{\min}$ & Avg ${\deg}_{\max}$ & Min & Max & Avg Density \\ \hline
2,000 & 5,000 & 3,495.1 & 2,049 & 66,355 & 12,115.3 & 2 & 968 & 485.3 & 1.3 & 20.8 & 4e-4 & 1e-2 & 2e-3 \\ \hline
\end{tabular}
\vspace{-0.3cm}
\end{table*}

Our settings of the DC-SBM generator are summarized in Table~\ref{Tab:Syn-Gen}, where `\textit{I}($a$, $b$)' and `\textit{F}($a$, $b$)' denote an integer and a float uniformly sampled from the range $[a, b]$; $T$ is the number of synthetic graphs; $N$ and $K$ are the numbers of nodes and communities in each graph; ${\deg}_{\min}$ and ${\deg}_{\max}$ are the minimum and maximum node degrees of a graph; $\gamma$ is a parameter to control the power-law distributions of node degrees with ${\deg} (v_i) \sim \{ {k^\gamma }/\sum\nolimits_{l = {{\deg}_{\min }}}^{{{\deg}_{\max }}} {{l^\gamma }} \}$ (${\deg}_{\min} \le k \le {\deg}_{\max}$); $\mu$ is a ratio between the number of within-community and between-community edges; $\rho$ is a parameter to control the heterogeneity of community size, with the size of each community following ${\mathop{\rm Dirichlet}\nolimits} (10/\rho )$.

\begin{algorithm}[t]\small
\caption{Generating a Synthetic Graph via DC-SBM}
\label{Alg:SBM}
\LinesNumbered
\KwIn{number of nodes $N_t$; number of communities $K_t$; range [${\deg}_{\min}, {\deg}_{\max}$] for node degrees; $\gamma$ for degree distribution; $\mu$ for numbers of within-community edges and between-community edges; $\rho$ for community size distribution}
\KwOut{a synthetic graph $\mathcal{G}_t = (\mathcal{V}_t, \mathcal{E}_t)$}
generate node set $\mathcal{V}_t = \{ v_1, \cdots, v_{N_t} \}$\\
\For{$r$ {\bf{from}} $1$ {\bf{to}} $K_t$}
{
    get size of the $r$-th community $|\mathcal{C}_r^{(t)}| \sim {\mathop{\rm Dirichlet}\nolimits} (10/\rho )$\\
    initialize $\mathcal{C}_r^{(t)}$'s degree sum $\varphi_r \leftarrow 0$
}
initialize total degree sum ${\bar \varphi} \leftarrow 0$\\
\For{{\bf{each}} node $v_i \in \mathcal{V}_t$}
{
    get $v_i$'s community label $c_i \sim {\mathop{\rm Multi}\nolimits} (|\mathcal{C}_1^{(t)}|, \cdots ,|\mathcal{C}_{{K_t}}^{(t)}|)$\\
    add $v_i$ into corresponding community $\mathcal{C}_{c_i}$\\
    get $v_i$'s degree ${\deg} (v_i) \sim \{ {k^\gamma }/\sum\limits_{l = {{\deg}_{\min }}}^{{{\deg}_{\max }}} {{l^\gamma }} |{{\deg}_{\min }} \le k \le {{\deg}_{\max }}\}$\\
    update degree sum $\varphi_{c_i} \leftarrow \varphi_{c_i} + {\deg} (v_i)$\\
    update total degree sum ${\bar \varphi} \leftarrow {\bar \varphi} + {\deg} (v_i)$
}
\For{{\bf{each}} node $v_i \in \mathcal{V}_t$}
{
    get $v_i$'s degree correction value ${\bf{\theta}}_i \leftarrow {\deg} (v_i)/\varphi_{c_i}$\\
}
\For{$r$ {\bf{from}} $1$ {\bf{to}} $K_t$}
{
    \For{$t$ {\bf{from}} $r$ {\bf{to}} $K_t$}
    {
        \If{$r = t$}
        {
            ${\bf{\Omega_{rr}}} \leftarrow \frac{\mu }{{1 + \mu }} \cdot {\varphi_r}$
        }
        \Else
        {
            ${\bf{\Omega_{rt}}} \leftarrow \frac{1}{{1 + \mu }} \cdot \frac{{{\varphi _r}{\varphi _t}}}{{\bar \varphi }}$
        }
    }
}
\For{$i$ {\bf{from}} $2$ {\bf{to}} $N_t$}
{
    \For{$j$ {\bf{from}} $1$ {\bf{to}} $i$}
    {
        generate an edge $(v_i, v_j)$ for $\mathcal{E}_t$ via ${\mathop{\rm Poisson}\nolimits} ({{\bf{\theta}}_i}{{\bf{\theta}}_j}{{\bf{\Omega }}_{{c_i}{c_j}}})$
    }
}
\end{algorithm}

For each synthetic graph $\mathcal{G}_t$, suppose there are $N_t$ nodes partitioned into $K_t$ communities, with $N_t$ and $K_t$ sampled from the corresponding distributions in Table~\ref{Tab:Syn-Gen}.
Let $\mathcal{C}^{(t)} = \{ \mathcal{C}_1^{(t)}, \cdots, \mathcal{C}_{K_t}^{(t)}\}$ denote the community ground-truth of $\mathcal{G}_t$.
We also sample $(\gamma, \mu, \rho)$ from corresponding distributions in Table~\ref{Tab:Syn-Gen}, which define the topology properties of $\mathcal{G}_t$. 
Algorithm~\ref{Alg:SBM} summarizes the procedure of generating a graph $\mathcal{G}_t$ via DC-SBM, based on the parameter settings of $\{ N_t, K_t, {\deg}_{\min}, {\deg}_{\max}, \gamma, \mu, \rho \}$. Concretely, DC-SBM generates an edge $(v_i, v_j)$ in each graph $\mathcal{G}_t$ via ${\bf{A}}^{(t)}_{ij} \sim {\mathop{\rm Poisson}\nolimits} ({\bf{\theta}}_i {\bf{\theta}}_j {\bf{\Omega}}_{{c_i}{c_j}})$, where $c_i$ is the community label of node $v_i$; ${\bf{\theta}} \in \mathbb{R}_+^{N_t}$ and ${\bf{\Omega}} \in \mathbb{R}_+^{K_t \times K_t}$ are the degree correction vector and block interaction matrix \cite{karrer2011stochastic} determined by $\{ \gamma, \mu, \rho \}$; ${\bf{\theta}}_i {\bf{\theta}}_j {\bf{\Omega}}_{{c_i}{c_j}}$ represents the expected number of edges between $v_i$ and $v_j$.

Statistics of the generated synthetic graphs (used for \textit{offline pre-training} in our experiments) are reported in Table~\ref{Tab:Stat-Syn-Data}.

\section{Detailed Complexity Analysis}\label{App-Cmp}

For each graph $\mathcal{G}'$, let $N$, $M$, and $d$ be the number of nodes, number of edges, and embedding dimensionality, respectively.
By using the efficient sparse-dense matrix multiplication, the complexity of the feature extraction defined in (\ref{Eq:Feat-Ext}) is $\mathcal{O}(Md + Nd^2)$. Similarly, the complexity of one FFP of the embedding encoder described by (\ref{Eq:Emb-Enc}) is no more than $\mathcal{O}(Md + Nd^2)$.

To derive a feasible CD result via Algorithm~\ref{Alg:Post-Prep}, the complexities of constructing the node pair set $\mathcal{P}$, one FFP of the \textit{binary classifier} to derive auxiliary graph $\mathcal{\tilde G}$, and extracting connected components via DFS/BFS are $\mathcal{O}(M + n_S)$, $\mathcal{O}((M + n_S)d^2)$, and $\mathcal{O}(N + {\tilde M})$, where $n_S$ is the number of randomly sampled node pairs; ${\tilde M} := |\mathcal{\tilde E}|$ is the number of edges in the auxiliary graph $\mathcal{\tilde G}$.

In summary, the complexity of \textit{online generalization} is $\mathcal{O}( (Md + Nd^2) + (Md + Nd^2) + (M + n_S) + (M + n_S)d^2 + (N + {\tilde M}) ) = \mathcal{O}((N + M)d^2)$, where we assume that $\tilde M \approx M$, and $n_S \ll M$.

\section{Detailed Experiment Setup}\label{App-Exp-Set}

\subsection{Datasets}\label{App-Data}

In the six real datasets (see Table~\ref{Tab:Data}), \textit{Protein}\footnote{https://downloads.thebiogrid.org/BioGRID} \cite{stark2006biogrid}
was collected based on the protein-protein interactions in the BioGRID repository.
\textit{ArXiv}\footnote{https://ogb.stanford.edu/docs/nodeprop/} \cite{wang2020microsoft}
and
\textit{DBLP}\footnote{https://snap.stanford.edu/data/com-DBLP.html} \cite{yang2012defining}
are public paper citation and collaboration graphs extracted from ArXiv and DBLP, respectively.
\textit{Amazon}\footnote{https://snap.stanford.edu/data/com-Amazon.html} \cite{yang2012defining}
was collected by crawling the product co-purchasing relations from Amazon, while
\textit{Youtube}\footnote{https://snap.stanford.edu/data/com-Youtube.html} \cite{yang2012defining}
was constructed based on the friendship relations of Youtube.
\textit{RoadCA}\footnote{https://snap.stanford.edu/data/roadNet-CA.html} \cite{leskovec2009community}
describes a road network in California.
To pre-process \textit{Protein}, we abstracted each protein as a unique node and constructed graph topology based on corresponding protein-protein interactions. Since there are multiple connected components in the extracted graph, we extracted the largest component for evaluation.
Moreover, we used original formats of the remaining datasets provided by their sources.

\subsection{Environments \& Implementation Details}
All the experiments were conducted on a server with one AMD EPYC 7742 64-Core CPU, $512$GB main memory, and one $80$GB memory GPU.
We used Python 3.7 to implement PRoCD, where the feature extraction module (\ref{Eq:Feat-Ext}), embedding encoder (\ref{Eq:Emb-Enc}), and binary node pair classifier (\ref{Eq:Bin-Clf}) were implemented via PyTorch 1.10.0. Hence, the feature extraction and FFP of PRoCD can be speeded up via the GPU. The efficient function `scipy.sparse.csgraph.connected\_components' was used to extract connected components in Algorithm~\ref{Alg:Post-Prep} to derive a feasible CD result.
For all the baselines, we adopted their official implementations and tuned parameters to report the best quality.
To ensure the fairness of comparison, the embedding dimensionality of all the embedding-based methods (i.e., \textit{LouvainNE}, \textit{SketchNE}, \textit{RaftGP}, \textit{ICD}, and \textbf{PRoCD}) was set to be the same (i.e., $64$).

\begin{table}[]\footnotesize
\caption{Detailed parameter settings of PRoCD.}\label{Tab:Param-Layer}
\begin{tabular}{l|lll|ccll}
\hline
\multirow{2}{*}{\textbf{Datasets}} & \multicolumn{3}{c|}{\textbf{Parameter Settings}} & \multicolumn{4}{c}{\textbf{Layer Configurations}} \\ \cline{2-8} 
 & \multicolumn{1}{l|}{($\alpha$, $\lambda$)} & \multicolumn{1}{l|}{($n_P$, $\eta$)} & $n_S$ & \multicolumn{1}{l|}{$d$} & \multicolumn{1}{l|}{$L_{\rm{Feat}}$} & \multicolumn{1}{l|}{$L_{\rm{GNN}}$} & $L_{\rm{BC}}$ \\ \hline
\textbf{Protein} & \multicolumn{1}{l|}{(0.1, 1)} & \multicolumn{1}{l|}{(20, 1e-4)} & 1e3 & \multicolumn{1}{c|}{\multirow{6}{*}{64}} & \multicolumn{1}{c|}{\multirow{6}{*}{2}} & \multicolumn{1}{l|}{4} & 2 \\
\textbf{ArXiv} & \multicolumn{1}{l|}{(1, 100)} & \multicolumn{1}{l|}{(100, 1e-4)} & 1e4 & \multicolumn{1}{c|}{} & \multicolumn{1}{c|}{} & \multicolumn{1}{l|}{2} & 6 \\
\textbf{DBLP} & \multicolumn{1}{l|}{(0.1, 10)} & \multicolumn{1}{l|}{(100, 1e-4)} & 1e4 & \multicolumn{1}{c|}{} & \multicolumn{1}{c|}{} & \multicolumn{1}{l|}{4} & 2 \\
\textbf{Amazon} & \multicolumn{1}{l|}{(0.1, 10)} & \multicolumn{1}{l|}{(100, 1e-4)} & 2e4 & \multicolumn{1}{c|}{} & \multicolumn{1}{c|}{} & \multicolumn{1}{l|}{4} & 2 \\
\textbf{Youtube} & \multicolumn{1}{l|}{(1e2, 1e2)} & \multicolumn{1}{l|}{(89, 1e-4)} & 1e4 & \multicolumn{1}{c|}{} & \multicolumn{1}{c|}{} & \multicolumn{1}{l|}{4} & 2 \\
\textbf{RoadCA} & \multicolumn{1}{l|}{(1e2, 10)} & \multicolumn{1}{l|}{(11, 1e-4)} & 1e4 & \multicolumn{1}{c|}{} & \multicolumn{1}{c|}{} & \multicolumn{1}{l|}{2} & 2 \\ \hline
\end{tabular}
\end{table}

\begin{figure}[]
 \begin{minipage}{0.49\linewidth}
 \subfigure[DBLP, w/ \textit{LPA}]{
  \includegraphics[width=\textwidth,trim=37 0 40 0,clip]{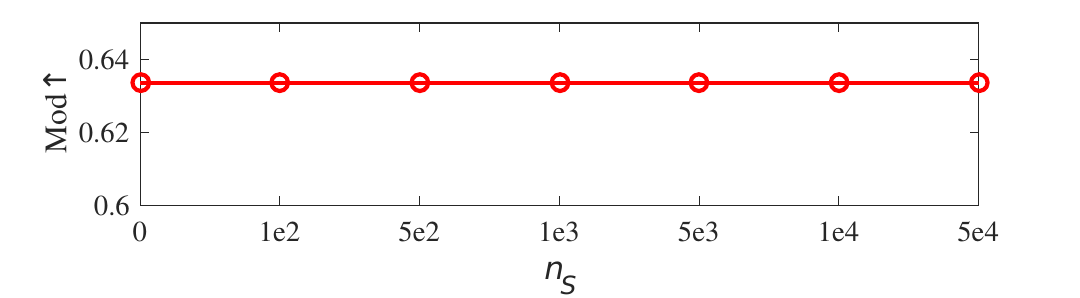}
  }
 \end{minipage}
 \begin{minipage}{0.49\linewidth}
 \subfigure[Amazon, w/ \textit{LPA}]{
  \includegraphics[width=\textwidth,trim=37 0 40 0,clip]{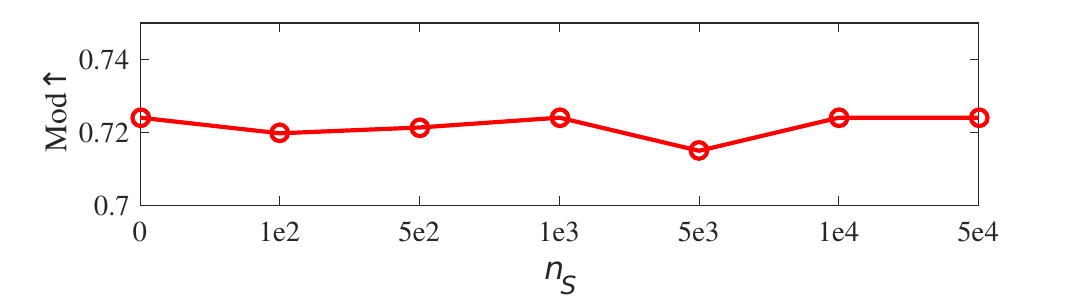}
  }
 \end{minipage}
 \begin{minipage}{0.49\linewidth}
 \subfigure[DBLP, w/ \textit{InfoMap}]{
  \includegraphics[width=\textwidth,trim=37 0 40 0,clip]{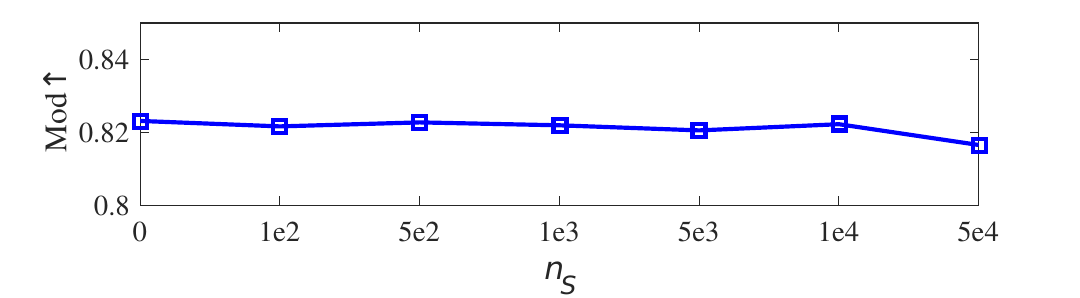}
  }
 \end{minipage}
 \begin{minipage}{0.49\linewidth}
 \subfigure[Amazon, w/ \textit{InfoMap}]{
  \includegraphics[width=\textwidth,trim=37 0 40 0,clip]{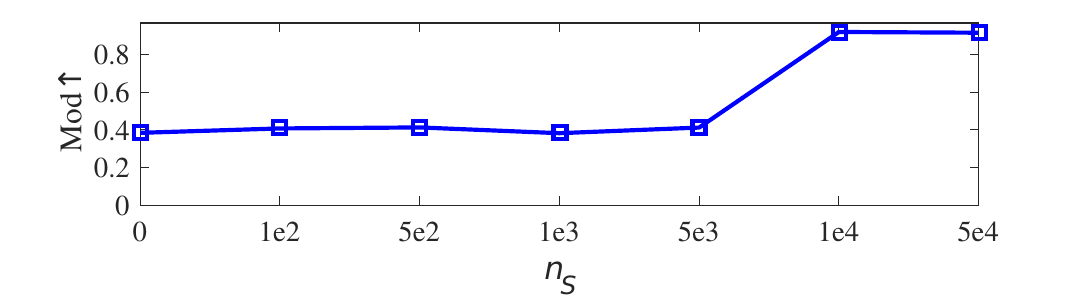}
  }
 \end{minipage}
 \begin{minipage}{0.49\linewidth}
 \subfigure[DBLP, w/ \textit{Locale}]{
  \includegraphics[width=\textwidth,trim=37 0 40 0,clip]{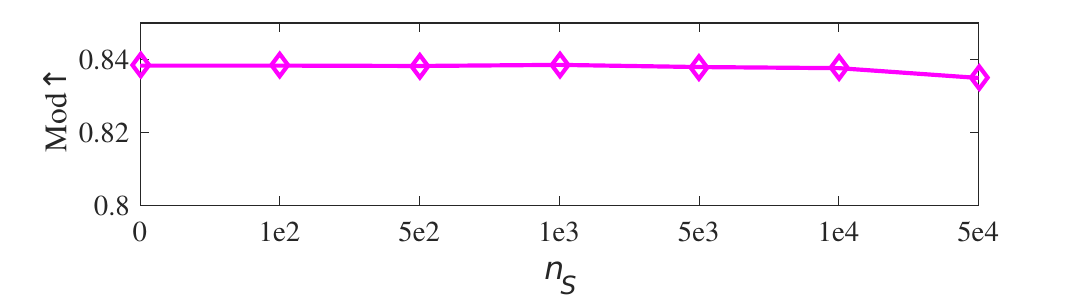}
  }
 \end{minipage}
 \begin{minipage}{0.49\linewidth}
 \subfigure[Amazon, w/ \textit{Locale}]{
  \includegraphics[width=\textwidth,trim=37 0 40 0,clip]{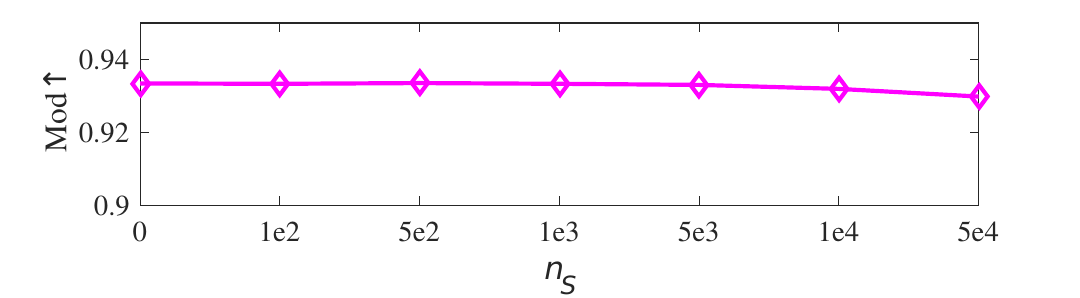}
  }
 \end{minipage}
 \caption{Parameter analysis w.r.t. the number of sampled node pairs $n_S$ on DBLP and Amazon in terms of modularity$\uparrow$.
 }\label{Fig:Abl-nS}
\end{figure}

\subsection{Parameter Settings \& Layer Configurations}\label{App-Param}

The recommended parameter settings and layer configurations of PRoCD are depicted in Table~\ref{Tab:Param-Layer}, where $\alpha$ and $\lambda$ are hyper-parameters in the pre-training objective (\ref{Eq:PTN-Obj}); $n_P$ and $\eta$ are the number of epochs and learning rate of pre-training in Algorithm~\ref{Alg:PTN}; $n_S$ is the number of randomly sampled node pairs in Algorithm~\ref{Alg:Post-Prep} for inference; $d$ is the embedding dimensionality; $L_{\rm{Feat}}$, $L_{\rm{GNN}}$, and $L_{\rm{BC}}$ are the numbers of MLP layers in (\ref{Eq:Feat-Ext}), GNN layers in (\ref{Eq:Emb-Enc}), and MLP layers in (\ref{Eq:Bin-Clf}).
Given an MLP, let ${\bf{u}}^{[s-1]}$ and ${\bf{u}}^{[s]}$ be the input and output of the $s$-th perceptron layer. The MLP in the feature extraction module (\ref{Eq:Feat-Ext}) is defined as
\begin{equation}
    {{\bf{u}}^{[s]}} = \tanh ({{\bf{u}}^{[s - 1]}}{{\bf{W}}^{[s]}} + {{\bf{b}}^{[s]}}),
\end{equation}
with $\{ {\bf{W}}^{[s]} \in \mathbb{R}^{d \times d}, {\bf{b}}^{[s]} \in \mathbb{R}^d \}$ as trainable parameters.
For the MLP $h_s(\cdot)$ (or $h_d(\cdot)$) in the binary classifier (\ref{Eq:Bin-Clf}), suppose there are $L$ layers.  The $s$-th layer ($s<L$) and the last layer of $h_s(\cdot)$ (or $h_d(\cdot)$) are defined as
\begin{equation}
    {{\bf{u}}^{[s]}} = \tanh ({{\bf{u}}^{[s - 1]}}{{\bf{W}}^{[s]}} + {{\bf{b}}^{[s]}}) + {{\bf{u}}^{[s - 1]}},
\end{equation}
\begin{equation}
    {{\bf{u}}^{[L]}} = {\mathop{\rm ReLU}\nolimits} ({{\bf{u}}^{[L - 1]}}{{\bf{W}}^{[L]}} + {{\bf{b}}^{[L]}}),
\end{equation}
where there is a skip connection in each perceptron layer except the last layer.

\section{Further Experiment Results}\label{App-Exp-Res}

We also conducted parameter analysis for the number of sampled node pairs (in Algorithm~\ref{Alg:Post-Prep}) for inference, where we set $n_S \in \{ 0, 1e2, 5e2, 1e3, 5e3, 1e4, 5e4 \}$.
Results on DBLP and Amazon in terms of modularity are shown in Fig.~\ref{Fig:Abl-nS}. In most cases except the variant with \textit{InfoMap} on Amazon, our PRoCD method is not sensitive to the setting of $n_S$. To ensure the inference quality of the exception case, one needs a large setting of $n_S$ (e.g., $n_S \ge 1e4$).

\section{Future Directions}\label{App:Future}

Some possible future directions are summarized as follows.

\textbf{Theoretical Analysis}.
This study empirically verified the potential of PRoCD to achieve a better trade-off between the quality and efficiency of $K$-agnostic.
However, most existing graph pre-training techniques lack theoretical guarantee for their transfer ability w.r.t. different pre-training data and downstream tasks.
We plan to theoretically analyze the transfer ability of PRoCD from small synthetic pre-training graphs to large real graphs, following previous analysis on random graph models (e.g., SBM \cite{qin2013regularized,jin2021improvements}).

\textbf{Extension to Dynamic Graphs}.
In this study, we considered $K$-agnostic CD on static graphs. CD on dynamic graphs \cite{rossetti2018community} is a more challenging setting, involving the variation of nodes, edges, and community membership over time. Usually, one can formulate a dynamic graph as a sequence of static snapshots with similar underlying properties \cite{lei2019gcn,qin2023temporal,qin2023high}. PRoCD can be easily extended to handle dynamic CD by directly generalizing the pre-trained model to each snapshot for fast online inference, without any retraining. Further validation of this extension is also our next focus.

\textbf{Integration of Graph Attributes}.
As stated in Section~\ref{Sec:Prob}, we considered CD without available graph attributes. A series of previous studies \cite{newman2016structure,qin2018adaptive,chunaev2020community,qin2021dual,zhao2022trade} have demonstrated the complicated correlations between graph topology and attributes, which are inherently heterogeneous information sources, for CD.
On the one hand, the integration of attributes may provide complementary characteristics for better CD quality. On the other hand, \textit{it may also incorporate noise or inconsistent features that lead to unexpected quality decline compared with those only considering one source}.
In our future work, we will explore the \textit{adaptive integration of attributes} for PRoCD. Concretely, when attributes match well with topology, we expect that PRoCD can fully utilize the complementary information of attributes to improve CD quality. \textit{When the two sources mismatch with one another, we try to adaptively control the contribution of attributes to avoid quality decline}.

\end{document}